\documentclass[aps,prl,reprint,groupedaddress]{revtex4-2}

\usepackage[colorlinks=true,citecolor=blue]{hyperref}

\usepackage{amsmath}
\usepackage{amssymb}
\usepackage{amsthm}
\usepackage{bbm}
\usepackage{verbatim}
\usepackage{graphicx}

\newcommand{\kb}{k_\mathrm{B}}

\newcommand{\Eint}{\int_0^\infty \! dE \;}

\newcommand{\ket}[1]{|#1\rangle}
\newcommand{\bra}[1]{\langle #1|}

\begin{document}


\title{Thermodynamic Uncertainty Relations for Coherent Transport}


\author{Kay Brandner\textsuperscript{1,2}, Keiji Saito\textsuperscript{3}}
\affiliation{
\textsuperscript{1}School of Physics and Astronomy, University of Nottingham, Nottingham NG7 2RD, United Kingdom\\
\textsuperscript{2}Centre for the Mathematics and Theoretical Physics of Quantum Non-Equilibrium Systems, University of Nottingham, Nottingham NG7 2RD, United Kingdom\\
\textsuperscript{3}Department of Physics, Kyoto University, Kyoto 606-8502, Japan
}


\date{\today}

\begin{abstract}
We derive a universal thermodynamic uncertainty relation for Fermionic coherent transport, which bounds the total rate of entropy production in terms of the mean and fluctuations of a single particle current. 
This bound holds for any multi-terminal geometry and arbitrary chemical and thermal biases, as long as no external magnetic fields are applied. 
It can further be saturated in two-terminal settings with boxcar-shaped transmission functions and reduces to its classical counterpart in linear response. 
Upon insertion of a numerical factor, our bound also extends to systems with broken time-reversal symmetry.
As an application, we derive trade-off relations between the figures of merit of coherent thermoelectric heat engines and refrigerators, which show that such devices can attain ideal efficiency only at vanishing mean power or diverging power fluctuations.
To illustrate our results, we work out a model of a coherent conductor consisting of a chain of quantum dots. 
\end{abstract}


\maketitle

Thermodynamic uncertainty relations have emerged as some of the most powerful tools in stochastic thermodynamics.
Going beyond the second law, they provide strictly positive lower bounds on the entropy production of thermodynamic processes in micro- and nanoscale systems, which are universal within well-defined settings, hold arbitrarily far from equilibrium, and depend solely on the mean and fluctuations of a single current \cite{seifert2018,seifert2019,horowitz2020}. 
These bounds have profound conceptual and practical implications. 
For instance, they give rise to trade-off relations between the figures of merit of steady-state heat engines and refrigerators, which show that such devices cannot reach ideal efficiency unless their power vanishes or the fluctuations of their input or output diverge \cite{pietzonka2018}. 
At the same time, thermodynamic uncertainty relations can be used to estimate the dissipation incurred by small-scale systems, even when only a fraction of their degrees of freedom is observable, a strategy known as thermodynamic inference \cite{seifert2018,seifert2019,horowitz2020}.
The prototype of all thermodynamic uncertainty relations,
\begin{equation}\label{eq:TURCl}
	\mathcal{Q}_\text{cl} = \frac{\sigma S}{2\kb J^2}\geq 1,
\end{equation}
applies to time-homogeneous Markov jump processes, where $\sigma$ is the total rate of entropy production, $J$ and $S$ are the mean and fluctuations of an arbitrary current and $\kb$ denotes Boltzmann's constant \cite{barato2015,gingrich2016}.
Since its discovery, wide-ranging generalizations and variations of this result have been derived in classical settings, including extensions to systems driven  by time-dependent control parameters \cite{koyuk2019,koyuk2020}, kinetic uncertainty relations \cite{garrahan2017,diterlizzi2019,yan2019, hiura2021} and fluctuation-response inequalities \cite{dechant2020, wang2020, chun2023, kwon2024a}. 
Beyond the classical realm, however, the picture is less complete. 
Although significant efforts were made to close this gap, see for instance Refs.~\cite{macieszczak2018, carollo2019, guarnieri2019, hasegawa2020, potanina2021, hasegawa2021, rignon-bret2021, menczel2021, vanvu2022, prech2023, vu2024, moreira2024, kwon2024}, quantum thermodynamic uncertainty relations with a similar degree of universality and practical relevance as their classical counterparts are still relatively scarce.

\begin{figure}[t!]
	\includegraphics[width=.48\textwidth]{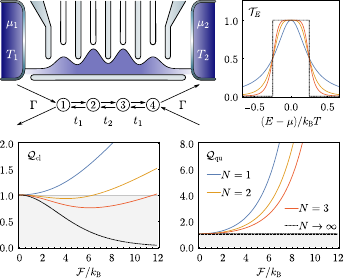}
	\caption{\label{Fig_1}
	\textbf{Top:} Chain of $N$ quantum dots in the coherent-tunneling regime.
		Carriers are exchanged between the reservoirs $1$ and $2$ through delocalized scattering states. 
		If the inter-dot hopping energies $t_j$ and the system-reservoir coupling $\Gamma$ are chosen according to Eq.~\eqref{eq:TunEng}, 
		the transmission function $\mathcal{T}_E$ is given by Eq.~\eqref{eq:TrasFExpl}. 
		This function is plotted here for $w=\kb T/2$ and approaches a boxcar shape in the limit $N\rightarrow\infty$. 
		\textbf{Bottom:} The classical uncertainty coefficient $\mathcal{Q}_\text{cl}$, defined in Eq.~\eqref{eq:TURCl}, becomes arbitrarily small for
		sufficiently large chemical biases $\mathcal{F}$ and numbers of dots $N$. 
		By contrast, the quantum uncertainty coefficient $\mathcal{Q}_\text{qu}$, defined in Eq.~\eqref{eq:TURQu}, satisfies $\mathcal{Q}_\text{qu}\geq 1$ for any $\mathcal{F}$
		and $N$.  
		For both plots, we choose the temperatures and chemical potentials of the reservoirs as $T_1 = T_2 = T$ and $\mu_1 = \mu + T\mathcal{F}/2$,
		$\mu_2 = \mu - T\mathcal{F}/2$.
	}
\end{figure}

Coherent conductors offer an appealing platform to revisit this problem. 
These systems consist of small samples exchanging transport carriers, which can be either actual particles or quasiparticle excitations, with multiple thermochemical reservoirs. 
If the mean free path of these carriers is sufficiently long, they traverse the sample without undergoing dephasing, inelastic collisions or interactions with each other, see Fig.~\ref{Fig_1}.  
Under these conditions, which can be realized in experiments with semiconductor nano-structures \cite{vanwees1988,wharam1988,schwab2000,matthews2014}, atomic junctions \cite{krans1995,vandenbrom1999,cui2017,lumbroso2018} or ultracold atomic gases \cite{brantut2012,brantut2013,krinner2015,lebrat2018}, the resulting transport process admits a simple and physically transparent description in terms of single-particle scattering amplitudes \cite{sivan1986,buttiker1992,benenti2017}. 
Using this formalism, it was shown early on that Eq.~\eqref{eq:TURCl} remains valid for Bosonic carriers \cite{saryal2019}. 
In Fermionic systems, however, a combination of energy filtering and Pauli-blocking can lead to an exponential suppression of current fluctuations in the chemical bias, and thus to arbitrary strong violations of the classical relation~\eqref{eq:TURCl} \cite{brandner2018,agarwalla2018,saryal2019,ehrlich2021,gerry2022,timpanaro2024}, see Fig.~\ref{Fig_1}. 
Following this observation, various relations between dissipation, current fluctuations and other quantities have been established for Fermionic coherent transport in the recent literature \cite{eriksson2021,tesser2023,tesser2024,acciai2024,palmqvist2024,blasi2025}. 
Still, the question whether there exists a universal thermodynamic uncertainty relation for this class of systems that involves the same quantities as its classical counterpart and provides practically relevant bounds on entropy production remains open so far. 
To help closing this gap, we here derive the relation 
\begin{equation}\label{eq:TURQu}
	\mathcal{Q}_\text{qu}=\frac{S_\alpha}{J_\alpha}
		\sinh\left[\frac{\sigma}{2\kb J_\alpha}\right]\geq 1,
\end{equation}
where $J_\alpha$ and $S_\alpha$ are the mean and fluctuations of the particle current that enters the sample from the reservoir $\alpha$, and $\sigma$ denotes the total rate of entropy production. 
This bound is our first key result. 
It holds for any geometry and potential landscape of the sample and arbitrary chemical and thermal biases, as long as the micro-dynamics of the system are symmetric under time reversal. 
It is further tight, since saturation can be achieved, at least in principle, in a two-terminal conductor with a narrow, boxcar-shaped transmission function, see Fig.~\ref{Fig_1}.

Four general remarks are in order before we move on to more technical aspects. 
First, we stress that, despite some formal similarity, Eq.~\eqref{eq:TURQu}, differs qualitatively from earlier thermodynamic uncertainty relations that were derived from fluctuation theorems \cite{hasegawa2019,timpanaro2019,potts2019,falasco2020}. 
The latter relations involve non-linear functions of an accumulated entropy production, which becomes arbitrary large at long times; they therefore provide meaningful constraints usually only for transient processes. 
By contrast, Eq.~\eqref{eq:TURQu} bounds the rate of entropy production in a steady state, which is independent of time.  
Second, while Eq.~\eqref{eq:TURQu} is derived here within the framework of coherent transport, it still applies in situations where incoherent or inelastic scattering events and interactions between carriers can be described effectively through probe terminals, i.e., virtual reservoirs whose intensive parameters are adjusted such that they do not exchange any heat or particles with the system on average \cite{buttiker1988}; 
the same approach can also be applied to mimic sources and sinks of energy inside the sample such as lattice vibrations in solid-state systems.
We therefore expect the bound \eqref{eq:TURQu} to be robust against moderate dephasing, internal dissipation and carrier interactions.
Third, close to equilibrium, the particle and heat currents entering the sample, $J_\alpha$ and $Q_\alpha$, become linear functions of the affinities $\mathcal{F}_\alpha = (\mu_\alpha-\mu)/T$ and $\mathcal{A}_\alpha = 1/T - 1/T_\alpha$, where $\mu_\alpha$ and $T_\alpha$ denote the chemical potential and temperature of the reservoir $\alpha$, and $\mu$ and $T$ are the corresponding reference values. 
The rate of entropy production, $\sigma = \sum_\alpha \mathcal{F}_\alpha J_\alpha + \mathcal{A}_\alpha Q_\alpha$, thus assumes a quadratic dependence on the affinities. 
Hence, if $\mathcal{F}_\alpha$ and $\mathcal{A}_\alpha$ are sufficiently small, the hyperbolic sine in Eq.~\eqref{eq:TURQu} can be replaced with its linear approximation. 
As a result, we recover the classical relation \eqref{eq:TURCl}. 
This observation shows that violations of the bound \eqref{eq:TURCl} can occur only beyond the linear-response regime. 
Finally, we note that the bound \eqref{eq:TURQu} becomes an equality for an isothermal two-terminal system with a narrow boxcar-shaped transmission function centered between the chemical potentials of the reservoirs, as was observed in Ref.~\cite{brandner2018}.

To derive the relation \eqref{eq:TURQu} and to discuss its implications quantitatively, we briefly review the scattering approach to coherent transport, where we focus on Fermionic systems throughout. 
The mean particle and heat currents that enter the sample through the terminal $\alpha$ are given by the Landauer-B\"uttiker formulas 
\begin{align}
	J_\alpha & = \frac{1}{h}\Eint \sum_\beta \mathcal{T}_E^{\alpha\beta}(f^\alpha_E - f^\beta_E), \\
	\label{eq:heatCurr}
	Q_\alpha & = \frac{1}{h}\Eint \sum_\beta \mathcal{T}_E^{\alpha\beta}(f^\alpha_E -f^\beta_E)(E-\mu_\alpha),
\end{align}
where $h$ denotes Planck's constant and
\begin{equation}
	f_E^\alpha = \frac{1}{1+\exp[(E-\mu_\alpha)/\kb T_\alpha]}
\end{equation}
is the Fermi function of the reservoir $\alpha$ \cite{sivan1986,buttiker1992,benenti2017}. 
The properties of the sample are encoded in the transmission functions and matrices   
\begin{equation}\label{eq:TMat}
	\mathcal{T}_E^{\alpha\beta} = \text{tr}[\mathsf{T}^{\alpha\beta}_E]
		\quad\text{and}\quad
		\mathsf{T}^{\alpha\beta}_E = \mathsf{S}^{\alpha\beta}_E (\mathsf{S}^{\alpha\beta}_E)^\dagger,
\end{equation}
where the scattering matrix $\mathsf{S}^{\alpha\beta}_E$ contains the amplitudes for coherent transitions of a single particle with energy $E$ from the reservoir $\beta$ to the reservoir $\alpha$; the trace in the formula for $\mathcal{T}^{\alpha\beta}_E$ indicates summation over all transport channels.
The fluctuations of the particle currents can be divided into thermal and a shot-noise contributions, $S_\alpha = S_\alpha^\text{th} + S_\alpha^\text{sh}$, which are given by  
\begin{align}
	\label{eq:thNoise}
	S_\alpha^\text{th} & = \frac{1}{h}\Eint \sum_{\beta\neq\alpha}
		\mathcal{T}^{\alpha\beta}_E (g^{\alpha\alpha}_E + g^{\beta\beta}_E),\\
	S_\alpha^\text{sh} & = \frac{1}{2h}\Eint \sum_{\beta\gamma}
		\text{tr}[\mathsf{T}^{\alpha\beta}_E \mathsf{T}^{\alpha\gamma}_E]
		(f^\beta_E - f^\gamma_E)^2
\end{align}
with $g^{\alpha\beta}_E = f^\alpha_E(1-f^\beta_E)$ \cite{sivan1986,buttiker1992,benenti2017}.
Since no dissipative processes occur inside a coherent conductor, entropy is generated only in the reservoirs. 
The total rate of entropy production thus becomes
\begin{equation}\label{eq:EntProd}
	\sigma = - \sum_\alpha Q_\alpha/T_\alpha.
\end{equation}

This expression can be rewritten in terms of transmission and Fermi functions by inserting the formula \eqref{eq:heatCurr}. 
Assuming that $\mathcal{T}^{\alpha\beta}_E = \mathcal{T}^{\beta\alpha}_E$,
it is then easy to show that
\begin{equation}\label{eq:locEntTRS}
	\sigma \geq \frac{\kb}{h}\Eint
		\sum_{\beta\neq\alpha}\mathcal{T}^{\alpha\beta}_E 
		(f^\alpha_E - f^\beta_E)
			\ln[g^{\alpha\beta}_E/g^{\beta\alpha}_E]
\end{equation}
for any given $\alpha$ \cite{brandner2025}.
The right-hand side of this inequality can be related to the thermal noise \eqref{eq:thNoise} as follows. 
We first introduce the new variables 
\begin{equation}
	X^{\alpha\beta}_E = (f^\alpha_E -f^\beta_E)/
		2[g^{\alpha\alpha}_E g^{\beta\beta}_E]^\frac{1}{2}
\end{equation}
and rewrite Eq.~\eqref{eq:locEntTRS} as  
\begin{equation}\label{eq:auxEntTRS}
	\sigma \geq \frac{4 \kb}{h} \Eint \sum_{\beta\neq\alpha}
		\mathcal{T}^{\alpha\beta}_E 
			[g^{\alpha\alpha}_E g^{\beta\beta}_E]^\frac{1}{2}
				\Phi[X^{\alpha\beta}_E]
\end{equation}
with $\Phi[x] = x\cdot \text{arsinh}[x] = x \cdot \ln[x+[1+x^2]^\frac{1}{2}]$. 
Since this function is convex, we have 
\begin{equation}
	\Phi[X^{\alpha\beta}_E] \geq \Phi[X] 
		+ \Phi'[X](X^{\alpha\beta}_E - X)
\end{equation}
for any real $X$. 
We now insert this bound into Eq.~\eqref{eq:auxEntTRS}, where we set 
\begin{equation}\label{eq:NDef}
	X	     = \frac{J_\alpha}{2 N_\alpha}, \quad
	N_\alpha = \frac{1}{h}\Eint \sum_{\beta\neq\alpha}
		\mathcal{T}^{\alpha\beta}_E
			[g^{\alpha\alpha}_E g^{\beta\beta}_E]^\frac{1}{2}
\end{equation}
such that the term proportional to the derivative $\Phi'$ of $\Phi$ vanishes. 
As a result, we obtain the inequality
\begin{equation}\label{eq:auxBndTRS}
	\sigma \geq  2 \kb J_\alpha \text{arsinh}[J_\alpha/ 2 N_\alpha].
\end{equation}
To eliminate $N_\alpha$, we note that $2N_\alpha \leq  S^\text{th}_\alpha \leq S_\alpha$,  where the first bound follows by comparing the Eqs.~\eqref{eq:thNoise} and \eqref{eq:NDef} and using the arithmetic-geometric mean inequality, and the second one by recalling that $S^\text{sh}_\alpha = S_\alpha-S^\text{th}_\alpha \geq 0$. 
Thus, since the inverse hyperbolic sine is antisymmetric and monotonically increasing, the bound \eqref{eq:auxBndTRS} remains valid after replacing $2N_\alpha$ with $S_\alpha$, which leads to Eq.~\eqref{eq:TURQu}. 

The above derivation uses the relation $\mathcal{T}^{\alpha\beta}_E = \mathcal{T}^{\beta\alpha}_E$, which holds in general only if the micro-dynamics of the system are symmetric under time reversal \cite{sivan1986,buttiker1992,benenti2017}. 
If this symmetry is broken, for instance due to an external magnetic field, the bound \eqref{eq:TURQu} is no longer applicable. 
As our second key result, we find that this restriction can be removed by introducing a numerical factor. 
Upon replacing the symmetry  $\mathcal{T}^{\alpha\beta}_E = \mathcal{T}^{\beta\alpha}_E$ with the sum rules $\sum_\beta \mathcal{T}^{\alpha\beta}_E = \sum_\beta \mathcal{T}^{\beta\alpha}_E$, which reflect the principle of current conservation and hold regardless of how the system behaves under time reversal \cite{sivan1986,buttiker1992,benenti2017}, we obtain the bound
\begin{equation}\label{eq:TURQuBTRS}
	\frac{S_\alpha}{J_\alpha}\sinh\biggl[
		\frac{\sigma}{\psi_0\kb J_\alpha}\biggr]\geq 1
\end{equation}
with $\psi_0 \simeq 0.85246 > 17/20$ \cite{nenciu2007,brandner2025}. 
This bound applies to any coherent conductor, including systems with chiral quantum Hall edge states, which are commonly used to engineer mesoscopic devices \cite{ji2003,granger2009,nam2013,jezouin2013,bauerle2018,sivre2019}.

\newcommand{\Th}{T_\text{h}}
\newcommand{\Tc}{T_\text{c}}

As one of their main applications, thermodynamic uncertainty relations make it possible to quantitatively describe the trade-off between different figures of merit of small-scale thermal machines \cite{pietzonka2018}. 
To show how such devices can be implemented with coherent conductors, we consider a two-terminal setup with $T_1 > T_2$ and $\mu_1 < \mu_2$. 
If $J_1>0$, this system acts as a thermoelectric heat engine with power output $P = (\mu_2-\mu_1)J_1$ and efficiency $\eta=P/Q_1\leq \eta_\text{C}$, where $\eta_\text{C}= 1- T_2/T_1$ denotes the Carnot bound. 
Alternatively, for $Q_2>0$, the system forms a thermoelectric refrigerator with output $Q=Q_2$ and efficiency, or coefficient of performance, $\varepsilon = - Q/P\leq \varepsilon_\text{C}$, where $\varepsilon_\text{C} = T_2/(T_1-T_2)$. 
In both cases, the output of the device and its thermodynamic efficiency are constrained by the relation \eqref{eq:TURQu}, which applies to any two-terminal conductor, since the sum rules $\sum_\beta \mathcal{T}^{\alpha\beta}_E = \sum_\beta \mathcal{T}^{\beta\alpha}_E$ imply $\mathcal{T}^{12}_E = \mathcal{T}^{21}_E =\mathcal{T}_E$ if $\alpha$ and $\beta$ take only two values.
Specifically, upon inserting the expression \eqref{eq:EntProd} for the rate of entropy production $\sigma$ and the first law $P= Q_1+Q_2$, we obtain the trade-off relations 
\begin{align}
	\label{eq:he}
	\mathcal{Q}^\text{HE}_\text{qu} = 
	\frac{S_P}{P(\mu_2-\mu_1)}\sinh\biggl[
		\frac{(\eta_\text{C}-\eta)(\mu_2-\mu_1)}{2\kb T_2\eta}\biggr]& 
			\geq 1,\\
	\label{eq:ref}
	\mathcal{Q}^\text{RF}_\text{qu} = 
	\frac{S_P\varepsilon}{Q(\mu_2-\mu_1)}\sinh\biggl[
		\frac{(\varepsilon_\text{C}-\varepsilon)(\mu_2-\mu_1)}{
			2\kb T_1 \varepsilon_\text{C}}\biggr] & \geq 1,
\end{align}
for heat engines and refrigerators, respectively, where the power fluctuations $S_P = (\mu_2-\mu_1)^2 S_1$ enter as a third figure of merit. 
These relations show, like their classical counterparts \cite{pietzonka2018}, 
\begin{align}
	\label{eq:clhe}
	\mathcal{Q}^\text{HE}_\text{cl} & = 
		\frac{1}{2\kb T_2}\frac{S_P(\eta_\text{C}-\eta)}{P\eta}\geq 1,\\
	\label{eq:clref}
	\mathcal{Q}^\text{RF}_\text{cl} & = 
		\frac{1}{2\kb T_1} 
		\frac{S_P\varepsilon(\varepsilon_\text{C} -\varepsilon)}	
			{Q\varepsilon_\text{C}}\geq 1,
\end{align}
which follow directly from the classical thermodynamic uncertainty relation \eqref{eq:TURCl}, that operation at ideal efficiency, i.e., $\eta = \eta_\text{C}$ or $\varepsilon=\varepsilon_\text{C}$, is possible only with either vanishing output or diverging power fluctuations. 
At the same time, the bounds \eqref{eq:he} and \eqref{eq:ref} can be rearranged as 
\begin{equation}\label{eq:to}
	\eta/\eta_\text{C} \leq 1/(1+T_2\mathcal{X})
		\quad\text{and}\quad
		\varepsilon/\varepsilon_\text{C} \leq 1 + T_1 \mathcal{X},
\end{equation}
where the parameter $\mathcal{X} = 2\kb\cdot\text{arsinh}[J_1/S_1]/(\mu_2 - \mu_1)$ depends only on the mean $J_1$ and fluctuations $S_1$ of the particle current, and the intensive properties of the reservoirs. 
The relations \eqref{eq:to} thus make it possible to estimate the efficiency of thermal machines without measuring any heat currents, which is often difficult in practice. 

\begin{figure}[t!]
	\includegraphics[width=.48\textwidth]{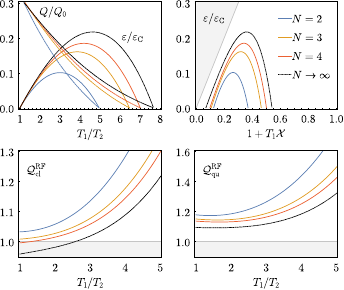}
	\caption{\label{Fig_2}
	Chain of quantum dots as a coherent refrigerator. 
	\textbf{Top left:}
	Solid lines show the normalized efficiency $\varepsilon/\varepsilon_\text{C}$ as a function of the temperature ratio $T_1/T_2$ for different numbers $N$ of dots; dashed lines indicate the corresponding heat current $Q$ in units of $Q_0=(\kb T_1)^2/10 h$.
	\textbf{Top right:}
	Normalized efficiency plotted against its estimated value from mean and fluctuations of the particle current $J_1$ for $T_1/T_2$ varying between $1$ and $8$. 
	The shaded area indicates the bound \eqref{eq:to}. 
	\textbf{Bottom:}
	The classical trade-off coefficient $\mathcal{Q}^\text{RF}_\text{cl}$, defined in Eq.~\eqref{eq:clref},
	falls below $1$ for $N\geq 4$ and sufficiently small temperature ratios, while its quantum counterpart 
	$\mathcal{Q}^\text{RF}_\text{qu}$, defined in Eq.~\eqref{eq:ref}, remains larger than $1$ throughout. 
	For all plots, we have set $\mu_1 = -2\kb T_1$, $\mu_2 = -\kb T_1/5$, $w=\kb T_1/2$ and varied $T_2$ while keeping $T_1$ fixed.
	}
\end{figure}

With these insights, which constitute our third main result, we conclude the general part of this article and move on to the specific setup of Fig.~\ref{Fig_1}. 
The key feature of this system is that its transmission function can be tuned to approximate a boxcar function. 
Such transmission profiles have been shown earlier to maximize the performance of coherent thermoelectric devices \cite{whitney2014,whitney2015,tesser2023a,mishra2024}, while at the same time leading to strong violations of the classical thermodynamic uncertainty relation \eqref{eq:TURCl} \cite{ehrlich2021,timpanaro2024}. 
To realize these effects in an experimentally relevant setting, chains of quantum dots provide a natural starting point \cite{volk2019,mills2019,mills2019a,qiao2020,hsiao2020,kandel2021,vandiepen2021}. 
The basic idea here is to generate a band-structured transmission spectrum with sharply bounded transmission windows through a spatially periodic scattering potential.
In addition, the couplings between the dots need to be fine-tuned to suppress oscillations of the transmission function, which appear generically for finite chains and can lead to large shot-noise  contributions enhancing the overall current fluctuations \cite{whitney2015,ehrlich2021}.

To find a suitable tunneling profile, we assume that the coupling strength $\Gamma$ between conductor and reservoirs and the internal hopping constants $t_j$ are energy independent. 
The transmission function of the chain is then given by the formula \cite{whitney2015}
\begin{equation}\label{eq:TransF}
	\mathcal{T}_E 
		= 4 \Gamma^2 \bigl|\bra{N}[E- H_\text{eff}]^{-1}\ket{1}\bigr|^2. 
\end{equation}
For a chain with nearest-neighbor hopping, the effective single-particle Hamiltonian takes the form
\begin{equation}
	H_\text{eff} = \sum_{j=1}^{N-1} t_j V_j
			- i\Gamma \ket{1}\bra{1} - i\Gamma \ket{N}\bra{N}
\end{equation}
with the tunneling operator $V_j = \ket{j}\bra{j+1}+ \ket{j+1}\bra{j}$ and $\ket{j}$ being the quantum state,  where the particle is localized at the dot $j$. 
With these prerequisites it is now in principle possible to optimize the parameters $\Gamma$ and $t_j$ with respect to some given objective function. 
Rather than delving into the intricacies of such optimization problems, which have been considered in Ref.~\cite{ehrlich2021}, we here make the ansatz 
\begin{align}
	\label{eq:TunEng}
	i\Gamma & = t_0 = t_N, \\
	t_j & = t_{N-j}  = 
	\frac{w}{4}\left[\sin\biggl[\frac{(2j+1)\pi}{2N}\biggr] 
		\sin\biggl[\frac{(2j-1)\pi}{2N}\biggr]\right]^{-\frac{1}{2}},\nonumber
\end{align}
which leads to \footnote{We have verified this result using computer algebra for $N=1,\dots,10$ and numerically for $N=11, \dots, 100$.}
\begin{equation}\label{eq:TrasFExpl}
	\mathcal{T}_E= \frac{1}{1 + (2 E/w)^{2N}}.
\end{equation}
This transmission function indeed converges to a boxcar function of width $w$ in the limit $N\rightarrow\infty$. 
The plots of Fig.~\ref{Fig_1} show that this approach is accompanied by increasingly strong violations of the classical bound \eqref{eq:TURCl}, while the quantum bound \eqref{eq:TURQu} is practically saturated over an increasingly large range of chemical biases. 
Fig.~\ref{Fig_2} illustrates the performance of the quantum dot chain as a thermoelectric refrigerator, where two main observations stand out. 
First, the bound \eqref{eq:to} provides a loose but non-trivial estimate of the efficiency of the device for a significant range of model parameters. 
Second, the classical trade-off relation \eqref{eq:clref} can be violated for sufficiently small temperature gradients and $N\geq 4$ dots. 
To understand why these violations are small, we recall that the temperatures of the reservoirs are encoded in the broadening of their Fermi functions. 
To generate a strong thermoelectric response, and thus positive cooling power, the chemical potentials must be tuned such that both Fermi edges overlap with the transmission window, which leads to significant thermal noise.
This situation differs qualitatively from the setting of Fig.~\ref{Fig_1}, where the Fermi energies of both reservoirs are far outside the transmission window and thermal noise is strongly suppressed as a result. 
Nonetheless, the fact that the classical trade-off relations \eqref{eq:clhe} and \eqref{eq:clref} can be violated shows that coherent thermal devices can offer a thermodynamic advantage over their classical counterparts. 
We leave it to future research to establish whether stronger violations of the classical relations \eqref{eq:clhe} and \eqref{eq:clref} can be achieved in more elaborate models and experimentally relevant systems, and whether their quantum counterparts \eqref{eq:he} and \eqref{eq:ref} can be strengthened by taking into account further practical constraints.

From a more general perspective, it remains to be seen whether the bounds \eqref{eq:TURQu} and \eqref{eq:TURQuBTRS} can be extended to energy or heat fluxes, which would require either structural changes or the incorporation of at least one additional parameter to account for the physical dimension of these quantities.  
It would further be interesting to explore whether the approach developed here can be adapted to coherent conductors that are subject to time-dependent driving fields \cite{potanina2021,brandner2020}, or involve superconducting junctions, which have recently been pointed out as an alternative means to overcome classical thermodynamic uncertainty relations \cite{ohnmacht2024}. 
Finally, we hope that the present work will encourage experimenters to test our theory and to prove that classical thermodynamic uncertainty relations can indeed be overcome with coherent conductors.

\begin{acknowledgments}
\emph{Acknowledgments.--}
This work was supported by the Medical Research Council (Grants No. MR/S034714/1 and MR/Y003845/1) and the Engineering and Physical Sciences Research Council (Grant No. EP/V031201/1).
K.S. was supported by the JSPS KAKENHI Grant No.~JP23K25796.
\end{acknowledgments}

\emph{Data availability.--} All data that support the findings of this study are included within the article and the Supplemental Material.


\begin{thebibliography}{81}%
\makeatletter
\providecommand \@ifxundefined [1]{%
 \@ifx{#1\undefined}
}%
\providecommand \@ifnum [1]{%
 \ifnum #1\expandafter \@firstoftwo
 \else \expandafter \@secondoftwo
 \fi
}%
\providecommand \@ifx [1]{%
 \ifx #1\expandafter \@firstoftwo
 \else \expandafter \@secondoftwo
 \fi
}%
\providecommand \natexlab [1]{#1}%
\providecommand \enquote  [1]{``#1''}%
\providecommand \bibnamefont  [1]{#1}%
\providecommand \bibfnamefont [1]{#1}%
\providecommand \citenamefont [1]{#1}%
\providecommand \href@noop [0]{\@secondoftwo}%
\providecommand \href [0]{\begingroup \@sanitize@url \@href}%
\providecommand \@href[1]{\@@startlink{#1}\@@href}%
\providecommand \@@href[1]{\endgroup#1\@@endlink}%
\providecommand \@sanitize@url [0]{\catcode `\\12\catcode `\$12\catcode
  `\&12\catcode `\#12\catcode `\^12\catcode `\_12\catcode `\%12\relax}%
\providecommand \@@startlink[1]{}%
\providecommand \@@endlink[0]{}%
\providecommand \url  [0]{\begingroup\@sanitize@url \@url }%
\providecommand \@url [1]{\endgroup\@href {#1}{\urlprefix }}%
\providecommand \urlprefix  [0]{URL }%
\providecommand \Eprint [0]{\href }%
\providecommand \doibase [0]{https://doi.org/}%
\providecommand \selectlanguage [0]{\@gobble}%
\providecommand \bibinfo  [0]{\@secondoftwo}%
\providecommand \bibfield  [0]{\@secondoftwo}%
\providecommand \translation [1]{[#1]}%
\providecommand \BibitemOpen [0]{}%
\providecommand \bibitemStop [0]{}%
\providecommand \bibitemNoStop [0]{.\EOS\space}%
\providecommand \EOS [0]{\spacefactor3000\relax}%
\providecommand \BibitemShut  [1]{\csname bibitem#1\endcsname}%
\let\auto@bib@innerbib\@empty
\bibitem [{\citenamefont {Seifert}(2018)}]{seifert2018}%
  \BibitemOpen
  \bibfield  {author} {\bibinfo {author} {\bibfnamefont {U.}~\bibnamefont
  {Seifert}},\ }\bibfield  {title} {\bibinfo {title} {Stochastic
  thermodynamics: {{From}} principles to the cost of precision},\ }\href
  {https://doi.org/10.1016/j.physa.2017.10.024} {\bibfield  {journal} {\bibinfo
   {journal} {Physica A}\ }\textbf {\bibinfo {volume} {504}},\ \bibinfo {pages}
  {176} (\bibinfo {year} {2018})}\BibitemShut {NoStop}%
\bibitem [{\citenamefont {Seifert}(2019)}]{seifert2019}%
  \BibitemOpen
  \bibfield  {author} {\bibinfo {author} {\bibfnamefont {U.}~\bibnamefont
  {Seifert}},\ }\bibfield  {title} {\bibinfo {title} {From {{Stochastic
  Thermodynamics}} to {{Thermodynamic Inference}}},\ }\href
  {https://doi.org/10.1146/annurev-conmatphys-031218-013554} {\bibfield
  {journal} {\bibinfo  {journal} {Annu. Rev. Condens. Matter Phys.}\ }\textbf
  {\bibinfo {volume} {10}},\ \bibinfo {pages} {171} (\bibinfo {year}
  {2019})}\BibitemShut {NoStop}%
\bibitem [{\citenamefont {Horowitz}\ and\ \citenamefont
  {Gingrich}(2020)}]{horowitz2020}%
  \BibitemOpen
  \bibfield  {author} {\bibinfo {author} {\bibfnamefont {J.~M.}\ \bibnamefont
  {Horowitz}}\ and\ \bibinfo {author} {\bibfnamefont {T.~R.}\ \bibnamefont
  {Gingrich}},\ }\bibfield  {title} {\bibinfo {title} {Thermodynamic
  uncertainty relations constrain non-equilibrium fluctuations},\ }\href
  {https://doi.org/10.1038/s41567-019-0702-6} {\bibfield  {journal} {\bibinfo
  {journal} {Nat. Phys.}\ }\textbf {\bibinfo {volume} {16}},\ \bibinfo {pages}
  {15} (\bibinfo {year} {2020})}\BibitemShut {NoStop}%
\bibitem [{\citenamefont {Pietzonka}\ and\ \citenamefont
  {Seifert}(2018)}]{pietzonka2018}%
  \BibitemOpen
  \bibfield  {author} {\bibinfo {author} {\bibfnamefont {P.}~\bibnamefont
  {Pietzonka}}\ and\ \bibinfo {author} {\bibfnamefont {U.}~\bibnamefont
  {Seifert}},\ }\bibfield  {title} {\bibinfo {title} {Universal {{Trade-Off}}
  between {{Power}}, {{Efficiency}}, and {{Constancy}} in {{Steady-State Heat
  Engines}}},\ }\href {https://doi.org/10.1103/PhysRevLett.120.190602}
  {\bibfield  {journal} {\bibinfo  {journal} {Phys. Rev. Lett.}\ }\textbf
  {\bibinfo {volume} {120}},\ \bibinfo {pages} {190602} (\bibinfo {year}
  {2018})}\BibitemShut {NoStop}%
\bibitem [{\citenamefont {Barato}\ and\ \citenamefont
  {Seifert}(2015)}]{barato2015}%
  \BibitemOpen
  \bibfield  {author} {\bibinfo {author} {\bibfnamefont {A.~C.}\ \bibnamefont
  {Barato}}\ and\ \bibinfo {author} {\bibfnamefont {U.}~\bibnamefont
  {Seifert}},\ }\bibfield  {title} {\bibinfo {title} {Thermodynamic
  {{Uncertainty Relation}} for {{Biomolecular Processes}}},\ }\href
  {https://doi.org/10.1103/PhysRevLett.114.158101} {\bibfield  {journal}
  {\bibinfo  {journal} {Phys. Rev. Lett.}\ }\textbf {\bibinfo {volume} {114}},\
  \bibinfo {pages} {158101} (\bibinfo {year} {2015})}\BibitemShut {NoStop}%
\bibitem [{\citenamefont {Gingrich}\ \emph {et~al.}(2016)\citenamefont
  {Gingrich}, \citenamefont {Horowitz}, \citenamefont {Perunov},\ and\
  \citenamefont {England}}]{gingrich2016}%
  \BibitemOpen
  \bibfield  {author} {\bibinfo {author} {\bibfnamefont {T.~R.}\ \bibnamefont
  {Gingrich}}, \bibinfo {author} {\bibfnamefont {J.~M.}\ \bibnamefont
  {Horowitz}}, \bibinfo {author} {\bibfnamefont {N.}~\bibnamefont {Perunov}},\
  and\ \bibinfo {author} {\bibfnamefont {J.~L.}\ \bibnamefont {England}},\
  }\bibfield  {title} {\bibinfo {title} {Dissipation {{Bounds All Steady-State
  Current Fluctuations}}},\ }\href
  {https://doi.org/10.1103/PhysRevLett.116.120601} {\bibfield  {journal}
  {\bibinfo  {journal} {Phys. Rev. Lett.}\ }\textbf {\bibinfo {volume} {116}},\
  \bibinfo {pages} {120601} (\bibinfo {year} {2016})}\BibitemShut {NoStop}%
\bibitem [{\citenamefont {Koyuk}\ and\ \citenamefont
  {Seifert}(2019)}]{koyuk2019}%
  \BibitemOpen
  \bibfield  {author} {\bibinfo {author} {\bibfnamefont {T.}~\bibnamefont
  {Koyuk}}\ and\ \bibinfo {author} {\bibfnamefont {U.}~\bibnamefont
  {Seifert}},\ }\bibfield  {title} {\bibinfo {title} {Operationally
  {{Accessible Bounds}} on {{Fluctuations}} and {{Entropy Production}} in
  {{Periodically Driven Systems}}},\ }\href
  {https://doi.org/10.1103/PhysRevLett.122.230601} {\bibfield  {journal}
  {\bibinfo  {journal} {Phys. Rev. Lett.}\ }\textbf {\bibinfo {volume} {122}},\
  \bibinfo {pages} {230601} (\bibinfo {year} {2019})}\BibitemShut {NoStop}%
\bibitem [{\citenamefont {Koyuk}\ and\ \citenamefont
  {Seifert}(2020)}]{koyuk2020}%
  \BibitemOpen
  \bibfield  {author} {\bibinfo {author} {\bibfnamefont {T.}~\bibnamefont
  {Koyuk}}\ and\ \bibinfo {author} {\bibfnamefont {U.}~\bibnamefont
  {Seifert}},\ }\bibfield  {title} {\bibinfo {title} {Thermodynamic
  {{Uncertainty Relation}} for {{Time-Dependent Driving}}},\ }\href
  {https://doi.org/10.1103/PhysRevLett.125.260604} {\bibfield  {journal}
  {\bibinfo  {journal} {Phys. Rev. Lett.}\ }\textbf {\bibinfo {volume} {125}},\
  \bibinfo {pages} {260604} (\bibinfo {year} {2020})}\BibitemShut {NoStop}%
\bibitem [{\citenamefont {Garrahan}(2017)}]{garrahan2017}%
  \BibitemOpen
  \bibfield  {author} {\bibinfo {author} {\bibfnamefont {J.~P.}\ \bibnamefont
  {Garrahan}},\ }\bibfield  {title} {\bibinfo {title} {Simple bounds on
  fluctuations and uncertainty relations for first-passage times of counting
  observables},\ }\href {https://doi.org/10.1103/PhysRevE.95.032134} {\bibfield
   {journal} {\bibinfo  {journal} {Phys. Rev. E}\ }\textbf {\bibinfo {volume}
  {95}},\ \bibinfo {pages} {032134} (\bibinfo {year} {2017})}\BibitemShut
  {NoStop}%
\bibitem [{\citenamefont {Di~Terlizzi}\ and\ \citenamefont
  {Baiesi}(2019)}]{diterlizzi2019}%
  \BibitemOpen
  \bibfield  {author} {\bibinfo {author} {\bibfnamefont {I.}~\bibnamefont
  {Di~Terlizzi}}\ and\ \bibinfo {author} {\bibfnamefont {M.}~\bibnamefont
  {Baiesi}},\ }\bibfield  {title} {\bibinfo {title} {Kinetic uncertainty
  relation},\ }\href {https://doi.org/10.1088/1751-8121/aaee34} {\bibfield
  {journal} {\bibinfo  {journal} {J. Phys. A: Math. Theor.}\ }\textbf {\bibinfo
  {volume} {52}},\ \bibinfo {pages} {02LT03} (\bibinfo {year}
  {2019})}\BibitemShut {NoStop}%
\bibitem [{\citenamefont {Yan}\ \emph {et~al.}(2019)\citenamefont {Yan},
  \citenamefont {Hilfinger}, \citenamefont {Vinnicombe},\ and\ \citenamefont
  {Paulsson}}]{yan2019}%
  \BibitemOpen
  \bibfield  {author} {\bibinfo {author} {\bibfnamefont {J.}~\bibnamefont
  {Yan}}, \bibinfo {author} {\bibfnamefont {A.}~\bibnamefont {Hilfinger}},
  \bibinfo {author} {\bibfnamefont {G.}~\bibnamefont {Vinnicombe}},\ and\
  \bibinfo {author} {\bibfnamefont {J.}~\bibnamefont {Paulsson}},\ }\bibfield
  {title} {\bibinfo {title} {Kinetic {{Uncertainty Relations}} for the
  {{Control}} of {{Stochastic Reaction Networks}}},\ }\href
  {https://doi.org/10.1103/PhysRevLett.123.108101} {\bibfield  {journal}
  {\bibinfo  {journal} {Phys. Rev. Lett.}\ }\textbf {\bibinfo {volume} {123}},\
  \bibinfo {pages} {108101} (\bibinfo {year} {2019})}\BibitemShut {NoStop}%
\bibitem [{\citenamefont {Hiura}\ and\ \citenamefont {Sasa}(2021)}]{hiura2021}%
  \BibitemOpen
  \bibfield  {author} {\bibinfo {author} {\bibfnamefont {K.}~\bibnamefont
  {Hiura}}\ and\ \bibinfo {author} {\bibfnamefont {S.-i.}\ \bibnamefont
  {Sasa}},\ }\bibfield  {title} {\bibinfo {title} {Kinetic uncertainty relation
  on first-passage time for accumulated current},\ }\href
  {https://doi.org/10.1103/PhysRevE.103.L050103} {\bibfield  {journal}
  {\bibinfo  {journal} {Phys. Rev. E}\ }\textbf {\bibinfo {volume} {103}},\
  \bibinfo {pages} {L050103} (\bibinfo {year} {2021})}\BibitemShut {NoStop}%
\bibitem [{\citenamefont {Dechant}\ and\ \citenamefont
  {Sasa}(2020)}]{dechant2020}%
  \BibitemOpen
  \bibfield  {author} {\bibinfo {author} {\bibfnamefont {A.}~\bibnamefont
  {Dechant}}\ and\ \bibinfo {author} {\bibfnamefont {S.-i.}\ \bibnamefont
  {Sasa}},\ }\bibfield  {title} {\bibinfo {title} {Fluctuation--response
  inequality out of equilibrium},\ }\href
  {https://doi.org/10.1073/pnas.1918386117} {\bibfield  {journal} {\bibinfo
  {journal} {Proc. Natl. Acad. Sci. U.S.A.}\ }\textbf {\bibinfo {volume}
  {117}},\ \bibinfo {pages} {6430} (\bibinfo {year} {2020})}\BibitemShut
  {NoStop}%
\bibitem [{\citenamefont {Wang}(2020)}]{wang2020}%
  \BibitemOpen
  \bibfield  {author} {\bibinfo {author} {\bibfnamefont {Y.}~\bibnamefont
  {Wang}},\ }\bibfield  {title} {\bibinfo {title} {Sub-{{Gaussian}} and
  subexponential fluctuation-response inequalities},\ }\href
  {https://doi.org/10.1103/PhysRevE.102.052105} {\bibfield  {journal} {\bibinfo
   {journal} {Phys. Rev. E}\ }\textbf {\bibinfo {volume} {102}},\ \bibinfo
  {pages} {052105} (\bibinfo {year} {2020})}\BibitemShut {NoStop}%
\bibitem [{\citenamefont {Chun}\ and\ \citenamefont
  {Horowitz}(2023)}]{chun2023}%
  \BibitemOpen
  \bibfield  {author} {\bibinfo {author} {\bibfnamefont {H.-M.}\ \bibnamefont
  {Chun}}\ and\ \bibinfo {author} {\bibfnamefont {J.~M.}\ \bibnamefont
  {Horowitz}},\ }\bibfield  {title} {\bibinfo {title} {Trade-offs between
  number fluctuations and response in nonequilibrium chemical reaction
  networks},\ }\href {https://doi.org/10.1063/5.0148662} {\bibfield  {journal}
  {\bibinfo  {journal} {The Journal of Chemical Physics}\ }\textbf {\bibinfo
  {volume} {158}},\ \bibinfo {pages} {174115} (\bibinfo {year}
  {2023})}\BibitemShut {NoStop}%
\bibitem [{\citenamefont {Kwon}\ \emph {et~al.}(2024)\citenamefont {Kwon},
  \citenamefont {Chun}, \citenamefont {Park},\ and\ \citenamefont
  {Lee}}]{kwon2024a}%
  \BibitemOpen
  \bibfield  {author} {\bibinfo {author} {\bibfnamefont {E.}~\bibnamefont
  {Kwon}}, \bibinfo {author} {\bibfnamefont {H.-M.}\ \bibnamefont {Chun}},
  \bibinfo {author} {\bibfnamefont {H.}~\bibnamefont {Park}},\ and\ \bibinfo
  {author} {\bibfnamefont {J.~S.}\ \bibnamefont {Lee}},\ }\href@noop {}
  {\bibinfo {title} {Fluctuation-response inequalities for kinetic and entropic
  perturbations}} (\bibinfo {year} {2024}),\ \Eprint
  {https://arxiv.org/abs/2411.18108} {arXiv:2411.18108} \BibitemShut {NoStop}%
\bibitem [{\citenamefont {Macieszczak}\ \emph {et~al.}(2018)\citenamefont
  {Macieszczak}, \citenamefont {Brandner},\ and\ \citenamefont
  {Garrahan}}]{macieszczak2018}%
  \BibitemOpen
  \bibfield  {author} {\bibinfo {author} {\bibfnamefont {K.}~\bibnamefont
  {Macieszczak}}, \bibinfo {author} {\bibfnamefont {K.}~\bibnamefont
  {Brandner}},\ and\ \bibinfo {author} {\bibfnamefont {J.~P.}\ \bibnamefont
  {Garrahan}},\ }\bibfield  {title} {\bibinfo {title} {Unified {{Thermodynamic
  Uncertainty Relations}} in {{Linear Response}}},\ }\href
  {https://doi.org/10.1103/PhysRevLett.121.130601} {\bibfield  {journal}
  {\bibinfo  {journal} {Phys. Rev. Lett.}\ }\textbf {\bibinfo {volume} {121}},\
  \bibinfo {pages} {130601} (\bibinfo {year} {2018})}\BibitemShut {NoStop}%
\bibitem [{\citenamefont {Carollo}\ \emph {et~al.}(2019)\citenamefont
  {Carollo}, \citenamefont {Jack},\ and\ \citenamefont
  {Garrahan}}]{carollo2019}%
  \BibitemOpen
  \bibfield  {author} {\bibinfo {author} {\bibfnamefont {F.}~\bibnamefont
  {Carollo}}, \bibinfo {author} {\bibfnamefont {R.~L.}\ \bibnamefont {Jack}},\
  and\ \bibinfo {author} {\bibfnamefont {J.~P.}\ \bibnamefont {Garrahan}},\
  }\bibfield  {title} {\bibinfo {title} {Unraveling the {{Large Deviation
  Statistics}} of {{Markovian Open Quantum Systems}}},\ }\href
  {https://doi.org/10.1103/PhysRevLett.122.130605} {\bibfield  {journal}
  {\bibinfo  {journal} {Phys. Rev. Lett.}\ }\textbf {\bibinfo {volume} {122}},\
  \bibinfo {pages} {130605} (\bibinfo {year} {2019})}\BibitemShut {NoStop}%
\bibitem [{\citenamefont {Guarnieri}\ \emph {et~al.}(2019)\citenamefont
  {Guarnieri}, \citenamefont {Landi}, \citenamefont {Clark},\ and\
  \citenamefont {Goold}}]{guarnieri2019}%
  \BibitemOpen
  \bibfield  {author} {\bibinfo {author} {\bibfnamefont {G.}~\bibnamefont
  {Guarnieri}}, \bibinfo {author} {\bibfnamefont {G.~T.}\ \bibnamefont
  {Landi}}, \bibinfo {author} {\bibfnamefont {S.~R.}\ \bibnamefont {Clark}},\
  and\ \bibinfo {author} {\bibfnamefont {J.}~\bibnamefont {Goold}},\ }\bibfield
   {title} {\bibinfo {title} {Thermodynamics of precision in quantum
  nonequilibrium steady states},\ }\href
  {https://doi.org/10.1103/PhysRevResearch.1.033021} {\bibfield  {journal}
  {\bibinfo  {journal} {Phys. Rev. Research}\ }\textbf {\bibinfo {volume}
  {1}},\ \bibinfo {pages} {033021} (\bibinfo {year} {2019})}\BibitemShut
  {NoStop}%
\bibitem [{\citenamefont {Hasegawa}(2020)}]{hasegawa2020}%
  \BibitemOpen
  \bibfield  {author} {\bibinfo {author} {\bibfnamefont {Y.}~\bibnamefont
  {Hasegawa}},\ }\bibfield  {title} {\bibinfo {title} {Quantum {{Thermodynamic
  Uncertainty Relation}} for {{Continuous Measurement}}},\ }\href
  {https://doi.org/10.1103/PhysRevLett.125.050601} {\bibfield  {journal}
  {\bibinfo  {journal} {Phys. Rev. Lett.}\ }\textbf {\bibinfo {volume} {125}},\
  \bibinfo {pages} {050601} (\bibinfo {year} {2020})}\BibitemShut {NoStop}%
\bibitem [{\citenamefont {Potanina}\ \emph {et~al.}(2021)\citenamefont
  {Potanina}, \citenamefont {Flindt}, \citenamefont {Moskalets},\ and\
  \citenamefont {Brandner}}]{potanina2021}%
  \BibitemOpen
  \bibfield  {author} {\bibinfo {author} {\bibfnamefont {E.}~\bibnamefont
  {Potanina}}, \bibinfo {author} {\bibfnamefont {C.}~\bibnamefont {Flindt}},
  \bibinfo {author} {\bibfnamefont {M.}~\bibnamefont {Moskalets}},\ and\
  \bibinfo {author} {\bibfnamefont {K.}~\bibnamefont {Brandner}},\ }\bibfield
  {title} {\bibinfo {title} {Thermodynamic bounds on coherent transport in
  periodically driven conductors},\ }\href
  {https://doi.org/10.1103/PhysRevX.11.021013} {\bibfield  {journal} {\bibinfo
  {journal} {Phys. Rev. X}\ }\textbf {\bibinfo {volume} {11}},\ \bibinfo
  {pages} {021013} (\bibinfo {year} {2021})}\BibitemShut {NoStop}%
\bibitem [{\citenamefont {Hasegawa}(2021)}]{hasegawa2021}%
  \BibitemOpen
  \bibfield  {author} {\bibinfo {author} {\bibfnamefont {Y.}~\bibnamefont
  {Hasegawa}},\ }\bibfield  {title} {\bibinfo {title} {Thermodynamic
  {{Uncertainty Relation}} for {{General Open Quantum Systems}}},\ }\href
  {https://doi.org/10.1103/PhysRevLett.126.010602} {\bibfield  {journal}
  {\bibinfo  {journal} {Phys. Rev. Lett.}\ }\textbf {\bibinfo {volume} {126}},\
  \bibinfo {pages} {010602} (\bibinfo {year} {2021})}\BibitemShut {NoStop}%
\bibitem [{\citenamefont {{Rignon-Bret}}\ \emph {et~al.}(2021)\citenamefont
  {{Rignon-Bret}}, \citenamefont {Guarnieri}, \citenamefont {Goold},\ and\
  \citenamefont {Mitchison}}]{rignon-bret2021}%
  \BibitemOpen
  \bibfield  {author} {\bibinfo {author} {\bibfnamefont {A.}~\bibnamefont
  {{Rignon-Bret}}}, \bibinfo {author} {\bibfnamefont {G.}~\bibnamefont
  {Guarnieri}}, \bibinfo {author} {\bibfnamefont {J.}~\bibnamefont {Goold}},\
  and\ \bibinfo {author} {\bibfnamefont {M.~T.}\ \bibnamefont {Mitchison}},\
  }\bibfield  {title} {\bibinfo {title} {Thermodynamics of precision in quantum
  nanomachines},\ }\href {https://doi.org/10.1103/PhysRevE.103.012133}
  {\bibfield  {journal} {\bibinfo  {journal} {Phys. Rev. E}\ }\textbf {\bibinfo
  {volume} {103}},\ \bibinfo {pages} {012133} (\bibinfo {year}
  {2021})}\BibitemShut {NoStop}%
\bibitem [{\citenamefont {Menczel}\ \emph {et~al.}(2021)\citenamefont
  {Menczel}, \citenamefont {Loisa}, \citenamefont {Brandner},\ and\
  \citenamefont {Flindt}}]{menczel2021}%
  \BibitemOpen
  \bibfield  {author} {\bibinfo {author} {\bibfnamefont {P.}~\bibnamefont
  {Menczel}}, \bibinfo {author} {\bibfnamefont {E.}~\bibnamefont {Loisa}},
  \bibinfo {author} {\bibfnamefont {K.}~\bibnamefont {Brandner}},\ and\
  \bibinfo {author} {\bibfnamefont {C.}~\bibnamefont {Flindt}},\ }\bibfield
  {title} {\bibinfo {title} {Thermodynamic uncertainty relations for coherently
  driven open quantum systems},\ }\href
  {https://doi.org/10.1088/1751-8121/ac0c8f} {\bibfield  {journal} {\bibinfo
  {journal} {J. Phys. A: Math. Theor.}\ }\textbf {\bibinfo {volume} {54}},\
  \bibinfo {pages} {314002} (\bibinfo {year} {2021})}\BibitemShut {NoStop}%
\bibitem [{\citenamefont {Van~Vu}\ and\ \citenamefont
  {Saito}(2022)}]{vanvu2022}%
  \BibitemOpen
  \bibfield  {author} {\bibinfo {author} {\bibfnamefont {T.}~\bibnamefont
  {Van~Vu}}\ and\ \bibinfo {author} {\bibfnamefont {K.}~\bibnamefont {Saito}},\
  }\bibfield  {title} {\bibinfo {title} {Thermodynamics of {{Precision}} in
  {{Markovian Open Quantum Dynamics}}},\ }\href
  {https://doi.org/10.1103/PhysRevLett.128.140602} {\bibfield  {journal}
  {\bibinfo  {journal} {Phys. Rev. Lett.}\ }\textbf {\bibinfo {volume} {128}},\
  \bibinfo {pages} {140602} (\bibinfo {year} {2022})}\BibitemShut {NoStop}%
\bibitem [{\citenamefont {Prech}\ \emph {et~al.}(2023)\citenamefont {Prech},
  \citenamefont {Johansson}, \citenamefont {Nyholm}, \citenamefont {Landi},
  \citenamefont {Verdozzi}, \citenamefont {Samuelsson},\ and\ \citenamefont
  {Potts}}]{prech2023}%
  \BibitemOpen
  \bibfield  {author} {\bibinfo {author} {\bibfnamefont {K.}~\bibnamefont
  {Prech}}, \bibinfo {author} {\bibfnamefont {P.}~\bibnamefont {Johansson}},
  \bibinfo {author} {\bibfnamefont {E.}~\bibnamefont {Nyholm}}, \bibinfo
  {author} {\bibfnamefont {G.~T.}\ \bibnamefont {Landi}}, \bibinfo {author}
  {\bibfnamefont {C.}~\bibnamefont {Verdozzi}}, \bibinfo {author}
  {\bibfnamefont {P.}~\bibnamefont {Samuelsson}},\ and\ \bibinfo {author}
  {\bibfnamefont {P.~P.}\ \bibnamefont {Potts}},\ }\bibfield  {title} {\bibinfo
  {title} {Entanglement and thermokinetic uncertainty relations in coherent
  mesoscopic transport},\ }\href
  {https://doi.org/10.1103/PhysRevResearch.5.023155} {\bibfield  {journal}
  {\bibinfo  {journal} {Phys. Rev. Research}\ }\textbf {\bibinfo {volume}
  {5}},\ \bibinfo {pages} {023155} (\bibinfo {year} {2023})}\BibitemShut
  {NoStop}%
\bibitem [{\citenamefont {Vu}(2024)}]{vu2024}%
  \BibitemOpen
  \bibfield  {author} {\bibinfo {author} {\bibfnamefont {T.~V.}\ \bibnamefont
  {Vu}},\ }\href@noop {} {\bibinfo {title} {Fundamental bounds on precision and
  response for quantum trajectory observables}} (\bibinfo {year} {2024}),\
  \Eprint {https://arxiv.org/abs/2411.19546} {arXiv:2411.19546} \BibitemShut
  {NoStop}%
\bibitem [{\citenamefont {Moreira}\ \emph {et~al.}(2024)\citenamefont
  {Moreira}, \citenamefont {Radaelli}, \citenamefont {Candeloro}, \citenamefont
  {Binder},\ and\ \citenamefont {Mitchison}}]{moreira2024}%
  \BibitemOpen
  \bibfield  {author} {\bibinfo {author} {\bibfnamefont {S.~V.}\ \bibnamefont
  {Moreira}}, \bibinfo {author} {\bibfnamefont {M.}~\bibnamefont {Radaelli}},
  \bibinfo {author} {\bibfnamefont {A.}~\bibnamefont {Candeloro}}, \bibinfo
  {author} {\bibfnamefont {F.~C.}\ \bibnamefont {Binder}},\ and\ \bibinfo
  {author} {\bibfnamefont {M.~T.}\ \bibnamefont {Mitchison}},\ }\href@noop {}
  {\bibinfo {title} {Precision bounds for multiple currents in open quantum
  systems}} (\bibinfo {year} {2024}),\ \Eprint
  {https://arxiv.org/abs/2411.09088} {arXiv:2411.09088} \BibitemShut {NoStop}%
\bibitem [{\citenamefont {Kwon}\ and\ \citenamefont {Lee}(2024)}]{kwon2024}%
  \BibitemOpen
  \bibfield  {author} {\bibinfo {author} {\bibfnamefont {E.}~\bibnamefont
  {Kwon}}\ and\ \bibinfo {author} {\bibfnamefont {J.~S.}\ \bibnamefont {Lee}},\
  }\href@noop {} {\bibinfo {title} {A unified framework for classical and
  quantum uncertainty relations using stochastic representations}} (\bibinfo
  {year} {2024}),\ \Eprint {https://arxiv.org/abs/2412.04988}
  {arXiv:2412.04988} \BibitemShut {NoStop}%
\bibitem [{\citenamefont {Van~Wees}\ \emph {et~al.}(1988)\citenamefont
  {Van~Wees}, \citenamefont {Van~Houten}, \citenamefont {Beenakker},
  \citenamefont {Williamson}, \citenamefont {Kouwenhoven}, \citenamefont {Van
  Der~Marel},\ and\ \citenamefont {Foxon}}]{vanwees1988}%
  \BibitemOpen
  \bibfield  {author} {\bibinfo {author} {\bibfnamefont {B.~J.}\ \bibnamefont
  {Van~Wees}}, \bibinfo {author} {\bibfnamefont {H.}~\bibnamefont
  {Van~Houten}}, \bibinfo {author} {\bibfnamefont {C.~W.~J.}\ \bibnamefont
  {Beenakker}}, \bibinfo {author} {\bibfnamefont {J.~G.}\ \bibnamefont
  {Williamson}}, \bibinfo {author} {\bibfnamefont {L.~P.}\ \bibnamefont
  {Kouwenhoven}}, \bibinfo {author} {\bibfnamefont {D.}~\bibnamefont {Van
  Der~Marel}},\ and\ \bibinfo {author} {\bibfnamefont {C.~T.}\ \bibnamefont
  {Foxon}},\ }\bibfield  {title} {\bibinfo {title} {Quantized conductance of
  point contacts in a two-dimensional electron gas},\ }\href
  {https://doi.org/10.1103/PhysRevLett.60.848} {\bibfield  {journal} {\bibinfo
  {journal} {Phys. Rev. Lett.}\ }\textbf {\bibinfo {volume} {60}},\ \bibinfo
  {pages} {848} (\bibinfo {year} {1988})}\BibitemShut {NoStop}%
\bibitem [{\citenamefont {Wharam}\ \emph {et~al.}(1988)\citenamefont {Wharam},
  \citenamefont {Thornton}, \citenamefont {Newbury}, \citenamefont {Pepper},
  \citenamefont {Ahmed}, \citenamefont {Frost}, \citenamefont {Hasko},
  \citenamefont {Peacock}, \citenamefont {Ritchie},\ and\ \citenamefont
  {Jones}}]{wharam1988}%
  \BibitemOpen
  \bibfield  {author} {\bibinfo {author} {\bibfnamefont {D.~A.}\ \bibnamefont
  {Wharam}}, \bibinfo {author} {\bibfnamefont {T.~J.}\ \bibnamefont
  {Thornton}}, \bibinfo {author} {\bibfnamefont {R.}~\bibnamefont {Newbury}},
  \bibinfo {author} {\bibfnamefont {M.}~\bibnamefont {Pepper}}, \bibinfo
  {author} {\bibfnamefont {H.}~\bibnamefont {Ahmed}}, \bibinfo {author}
  {\bibfnamefont {J.~E.~F.}\ \bibnamefont {Frost}}, \bibinfo {author}
  {\bibfnamefont {D.~G.}\ \bibnamefont {Hasko}}, \bibinfo {author}
  {\bibfnamefont {D.~C.}\ \bibnamefont {Peacock}}, \bibinfo {author}
  {\bibfnamefont {D.~A.}\ \bibnamefont {Ritchie}},\ and\ \bibinfo {author}
  {\bibfnamefont {G.~A.~C.}\ \bibnamefont {Jones}},\ }\bibfield  {title}
  {\bibinfo {title} {One-dimensional transport and the quantisation of the
  ballistic resistance},\ }\href {https://doi.org/10.1088/0022-3719/21/8/002}
  {\bibfield  {journal} {\bibinfo  {journal} {J. Phys. C: Solid State Phys.}\
  }\textbf {\bibinfo {volume} {21}},\ \bibinfo {pages} {L209} (\bibinfo {year}
  {1988})}\BibitemShut {NoStop}%
\bibitem [{\citenamefont {Schwab}\ \emph {et~al.}(2000)\citenamefont {Schwab},
  \citenamefont {Henriksen}, \citenamefont {Worlock},\ and\ \citenamefont
  {Roukes}}]{schwab2000}%
  \BibitemOpen
  \bibfield  {author} {\bibinfo {author} {\bibfnamefont {K.}~\bibnamefont
  {Schwab}}, \bibinfo {author} {\bibfnamefont {E.~A.}\ \bibnamefont
  {Henriksen}}, \bibinfo {author} {\bibfnamefont {J.~M.}\ \bibnamefont
  {Worlock}},\ and\ \bibinfo {author} {\bibfnamefont {M.~L.}\ \bibnamefont
  {Roukes}},\ }\bibfield  {title} {\bibinfo {title} {Measurement of the quantum
  of thermal conductance},\ }\href {https://doi.org/10.1038/35010065}
  {\bibfield  {journal} {\bibinfo  {journal} {Nature}\ }\textbf {\bibinfo
  {volume} {404}},\ \bibinfo {pages} {974} (\bibinfo {year}
  {2000})}\BibitemShut {NoStop}%
\bibitem [{\citenamefont {Matthews}\ \emph {et~al.}(2014)\citenamefont
  {Matthews}, \citenamefont {Battista}, \citenamefont {S{\'a}nchez},
  \citenamefont {Samuelsson},\ and\ \citenamefont {Linke}}]{matthews2014}%
  \BibitemOpen
  \bibfield  {author} {\bibinfo {author} {\bibfnamefont {J.}~\bibnamefont
  {Matthews}}, \bibinfo {author} {\bibfnamefont {F.}~\bibnamefont {Battista}},
  \bibinfo {author} {\bibfnamefont {D.}~\bibnamefont {S{\'a}nchez}}, \bibinfo
  {author} {\bibfnamefont {P.}~\bibnamefont {Samuelsson}},\ and\ \bibinfo
  {author} {\bibfnamefont {H.}~\bibnamefont {Linke}},\ }\bibfield  {title}
  {\bibinfo {title} {Experimental verification of reciprocity relations in
  quantum thermoelectric transport},\ }\href
  {https://doi.org/10.1103/PhysRevB.90.165428} {\bibfield  {journal} {\bibinfo
  {journal} {Phys. Rev. B}\ }\textbf {\bibinfo {volume} {90}},\ \bibinfo
  {pages} {165428} (\bibinfo {year} {2014})}\BibitemShut {NoStop}%
\bibitem [{\citenamefont {Krans}\ \emph {et~al.}(1995)\citenamefont {Krans},
  \citenamefont {Van~Ruitenbeek}, \citenamefont {Fisun}, \citenamefont
  {Yanson},\ and\ \citenamefont {De~Jongh}}]{krans1995}%
  \BibitemOpen
  \bibfield  {author} {\bibinfo {author} {\bibfnamefont {J.~M.}\ \bibnamefont
  {Krans}}, \bibinfo {author} {\bibfnamefont {J.~M.}\ \bibnamefont
  {Van~Ruitenbeek}}, \bibinfo {author} {\bibfnamefont {V.~V.}\ \bibnamefont
  {Fisun}}, \bibinfo {author} {\bibfnamefont {I.~K.}\ \bibnamefont {Yanson}},\
  and\ \bibinfo {author} {\bibfnamefont {L.~J.}\ \bibnamefont {De~Jongh}},\
  }\bibfield  {title} {\bibinfo {title} {The signature of conductance
  quantization in metallic point contacts},\ }\href
  {https://doi.org/10.1038/375767a0} {\bibfield  {journal} {\bibinfo  {journal}
  {Nature}\ }\textbf {\bibinfo {volume} {375}},\ \bibinfo {pages} {767}
  (\bibinfo {year} {1995})}\BibitemShut {NoStop}%
\bibitem [{\citenamefont {Van Den~Brom}\ and\ \citenamefont
  {Van~Ruitenbeek}(1999)}]{vandenbrom1999}%
  \BibitemOpen
  \bibfield  {author} {\bibinfo {author} {\bibfnamefont {H.~E.}\ \bibnamefont
  {Van Den~Brom}}\ and\ \bibinfo {author} {\bibfnamefont {J.~M.}\ \bibnamefont
  {Van~Ruitenbeek}},\ }\bibfield  {title} {\bibinfo {title} {Quantum
  {{Suppression}} of {{Shot Noise}} in {{Atom-Size Metallic Contacts}}},\
  }\href {https://doi.org/10.1103/PhysRevLett.82.1526} {\bibfield  {journal}
  {\bibinfo  {journal} {Phys. Rev. Lett.}\ }\textbf {\bibinfo {volume} {82}},\
  \bibinfo {pages} {1526} (\bibinfo {year} {1999})}\BibitemShut {NoStop}%
\bibitem [{\citenamefont {Cui}\ \emph {et~al.}(2017)\citenamefont {Cui},
  \citenamefont {Jeong}, \citenamefont {Hur}, \citenamefont {Matt},
  \citenamefont {Kl{\"o}ckner}, \citenamefont {Pauly}, \citenamefont {Nielaba},
  \citenamefont {Cuevas}, \citenamefont {Meyhofer},\ and\ \citenamefont
  {Reddy}}]{cui2017}%
  \BibitemOpen
  \bibfield  {author} {\bibinfo {author} {\bibfnamefont {L.}~\bibnamefont
  {Cui}}, \bibinfo {author} {\bibfnamefont {W.}~\bibnamefont {Jeong}}, \bibinfo
  {author} {\bibfnamefont {S.}~\bibnamefont {Hur}}, \bibinfo {author}
  {\bibfnamefont {M.}~\bibnamefont {Matt}}, \bibinfo {author} {\bibfnamefont
  {J.~C.}\ \bibnamefont {Kl{\"o}ckner}}, \bibinfo {author} {\bibfnamefont
  {F.}~\bibnamefont {Pauly}}, \bibinfo {author} {\bibfnamefont
  {P.}~\bibnamefont {Nielaba}}, \bibinfo {author} {\bibfnamefont {J.~C.}\
  \bibnamefont {Cuevas}}, \bibinfo {author} {\bibfnamefont {E.}~\bibnamefont
  {Meyhofer}},\ and\ \bibinfo {author} {\bibfnamefont {P.}~\bibnamefont
  {Reddy}},\ }\bibfield  {title} {\bibinfo {title} {Quantized thermal transport
  in single-atom junctions},\ }\href {https://doi.org/10.1126/science.aam6622}
  {\bibfield  {journal} {\bibinfo  {journal} {Science}\ }\textbf {\bibinfo
  {volume} {355}},\ \bibinfo {pages} {1192} (\bibinfo {year}
  {2017})}\BibitemShut {NoStop}%
\bibitem [{\citenamefont {Lumbroso}\ \emph {et~al.}(2018)\citenamefont
  {Lumbroso}, \citenamefont {Simine}, \citenamefont {Nitzan}, \citenamefont
  {Segal},\ and\ \citenamefont {Tal}}]{lumbroso2018}%
  \BibitemOpen
  \bibfield  {author} {\bibinfo {author} {\bibfnamefont {O.~S.}\ \bibnamefont
  {Lumbroso}}, \bibinfo {author} {\bibfnamefont {L.}~\bibnamefont {Simine}},
  \bibinfo {author} {\bibfnamefont {A.}~\bibnamefont {Nitzan}}, \bibinfo
  {author} {\bibfnamefont {D.}~\bibnamefont {Segal}},\ and\ \bibinfo {author}
  {\bibfnamefont {O.}~\bibnamefont {Tal}},\ }\bibfield  {title} {\bibinfo
  {title} {Electronic noise due to temperature differences in atomic-scale
  junctions},\ }\href {https://doi.org/10.1038/s41586-018-0592-2} {\bibfield
  {journal} {\bibinfo  {journal} {Nature}\ }\textbf {\bibinfo {volume} {562}},\
  \bibinfo {pages} {240} (\bibinfo {year} {2018})}\BibitemShut {NoStop}%
\bibitem [{\citenamefont {Brantut}\ \emph {et~al.}(2012)\citenamefont
  {Brantut}, \citenamefont {Meineke}, \citenamefont {Stadler}, \citenamefont
  {Krinner},\ and\ \citenamefont {Esslinger}}]{brantut2012}%
  \BibitemOpen
  \bibfield  {author} {\bibinfo {author} {\bibfnamefont {J.-P.}\ \bibnamefont
  {Brantut}}, \bibinfo {author} {\bibfnamefont {J.}~\bibnamefont {Meineke}},
  \bibinfo {author} {\bibfnamefont {D.}~\bibnamefont {Stadler}}, \bibinfo
  {author} {\bibfnamefont {S.}~\bibnamefont {Krinner}},\ and\ \bibinfo {author}
  {\bibfnamefont {T.}~\bibnamefont {Esslinger}},\ }\bibfield  {title} {\bibinfo
  {title} {Conduction of {{Ultracold Fermions Through}} a {{Mesoscopic
  Channel}}},\ }\href {https://doi.org/10.1126/science.1223175} {\bibfield
  {journal} {\bibinfo  {journal} {Science}\ }\textbf {\bibinfo {volume}
  {337}},\ \bibinfo {pages} {1069} (\bibinfo {year} {2012})}\BibitemShut
  {NoStop}%
\bibitem [{\citenamefont {Brantut}\ \emph {et~al.}(2013)\citenamefont
  {Brantut}, \citenamefont {Grenier}, \citenamefont {Meineke}, \citenamefont
  {Stadler}, \citenamefont {Krinner}, \citenamefont {Kollath}, \citenamefont
  {Esslinger},\ and\ \citenamefont {Georges}}]{brantut2013}%
  \BibitemOpen
  \bibfield  {author} {\bibinfo {author} {\bibfnamefont {J.-P.}\ \bibnamefont
  {Brantut}}, \bibinfo {author} {\bibfnamefont {C.}~\bibnamefont {Grenier}},
  \bibinfo {author} {\bibfnamefont {J.}~\bibnamefont {Meineke}}, \bibinfo
  {author} {\bibfnamefont {D.}~\bibnamefont {Stadler}}, \bibinfo {author}
  {\bibfnamefont {S.}~\bibnamefont {Krinner}}, \bibinfo {author} {\bibfnamefont
  {C.}~\bibnamefont {Kollath}}, \bibinfo {author} {\bibfnamefont
  {T.}~\bibnamefont {Esslinger}},\ and\ \bibinfo {author} {\bibfnamefont
  {A.}~\bibnamefont {Georges}},\ }\bibfield  {title} {\bibinfo {title} {A
  {{Thermoelectric Heat Engine}} with {{Ultracold Atoms}}},\ }\href
  {https://doi.org/10.1126/science.1242308} {\bibfield  {journal} {\bibinfo
  {journal} {Science}\ }\textbf {\bibinfo {volume} {342}},\ \bibinfo {pages}
  {713} (\bibinfo {year} {2013})}\BibitemShut {NoStop}%
\bibitem [{\citenamefont {Krinner}\ \emph {et~al.}(2015)\citenamefont
  {Krinner}, \citenamefont {Stadler}, \citenamefont {Husmann}, \citenamefont
  {Brantut},\ and\ \citenamefont {Esslinger}}]{krinner2015}%
  \BibitemOpen
  \bibfield  {author} {\bibinfo {author} {\bibfnamefont {S.}~\bibnamefont
  {Krinner}}, \bibinfo {author} {\bibfnamefont {D.}~\bibnamefont {Stadler}},
  \bibinfo {author} {\bibfnamefont {D.}~\bibnamefont {Husmann}}, \bibinfo
  {author} {\bibfnamefont {J.-P.}\ \bibnamefont {Brantut}},\ and\ \bibinfo
  {author} {\bibfnamefont {T.}~\bibnamefont {Esslinger}},\ }\bibfield  {title}
  {\bibinfo {title} {Observation of quantized conductance in neutral matter},\
  }\href {https://doi.org/10.1038/nature14049} {\bibfield  {journal} {\bibinfo
  {journal} {Nature}\ }\textbf {\bibinfo {volume} {517}},\ \bibinfo {pages}
  {64} (\bibinfo {year} {2015})}\BibitemShut {NoStop}%
\bibitem [{\citenamefont {Lebrat}\ \emph {et~al.}(2018)\citenamefont {Lebrat},
  \citenamefont {Gri{\v s}ins}, \citenamefont {Husmann}, \citenamefont
  {H{\"a}usler}, \citenamefont {Corman}, \citenamefont {Giamarchi},
  \citenamefont {Brantut},\ and\ \citenamefont {Esslinger}}]{lebrat2018}%
  \BibitemOpen
  \bibfield  {author} {\bibinfo {author} {\bibfnamefont {M.}~\bibnamefont
  {Lebrat}}, \bibinfo {author} {\bibfnamefont {P.}~\bibnamefont {Gri{\v
  s}ins}}, \bibinfo {author} {\bibfnamefont {D.}~\bibnamefont {Husmann}},
  \bibinfo {author} {\bibfnamefont {S.}~\bibnamefont {H{\"a}usler}}, \bibinfo
  {author} {\bibfnamefont {L.}~\bibnamefont {Corman}}, \bibinfo {author}
  {\bibfnamefont {T.}~\bibnamefont {Giamarchi}}, \bibinfo {author}
  {\bibfnamefont {J.-P.}\ \bibnamefont {Brantut}},\ and\ \bibinfo {author}
  {\bibfnamefont {T.}~\bibnamefont {Esslinger}},\ }\bibfield  {title} {\bibinfo
  {title} {Band and {{Correlated Insulators}} of {{Cold Fermions}} in a
  {{Mesoscopic Lattice}}},\ }\href {https://doi.org/10.1103/PhysRevX.8.011053}
  {\bibfield  {journal} {\bibinfo  {journal} {Phys. Rev. X}\ }\textbf {\bibinfo
  {volume} {8}},\ \bibinfo {pages} {011053} (\bibinfo {year}
  {2018})}\BibitemShut {NoStop}%
\bibitem [{\citenamefont {Sivan}\ and\ \citenamefont {Imry}(1986)}]{sivan1986}%
  \BibitemOpen
  \bibfield  {author} {\bibinfo {author} {\bibfnamefont {U.}~\bibnamefont
  {Sivan}}\ and\ \bibinfo {author} {\bibfnamefont {Y.}~\bibnamefont {Imry}},\
  }\bibfield  {title} {\bibinfo {title} {Multichannel {{Landauer}} formula for
  thermoelectric transport with application to thermopower near the mobility
  edge},\ }\href {https://doi.org/10.1103/PhysRevB.33.551} {\bibfield
  {journal} {\bibinfo  {journal} {Phys. Rev. B}\ }\textbf {\bibinfo {volume}
  {33}},\ \bibinfo {pages} {551} (\bibinfo {year} {1986})}\BibitemShut
  {NoStop}%
\bibitem [{\citenamefont {B{\"u}ttiker}(1992)}]{buttiker1992}%
  \BibitemOpen
  \bibfield  {author} {\bibinfo {author} {\bibfnamefont {M.}~\bibnamefont
  {B{\"u}ttiker}},\ }\bibfield  {title} {\bibinfo {title} {Scattering theory of
  current and intensity noise correlations in conductors and wave guides},\
  }\href {https://doi.org/10.1103/PhysRevB.46.12485} {\bibfield  {journal}
  {\bibinfo  {journal} {Phys. Rev. B}\ }\textbf {\bibinfo {volume} {46}},\
  \bibinfo {pages} {12485} (\bibinfo {year} {1992})}\BibitemShut {NoStop}%
\bibitem [{\citenamefont {Benenti}\ \emph {et~al.}(2017)\citenamefont
  {Benenti}, \citenamefont {Casati}, \citenamefont {Saito},\ and\ \citenamefont
  {Whitney}}]{benenti2017}%
  \BibitemOpen
  \bibfield  {author} {\bibinfo {author} {\bibfnamefont {G.}~\bibnamefont
  {Benenti}}, \bibinfo {author} {\bibfnamefont {G.}~\bibnamefont {Casati}},
  \bibinfo {author} {\bibfnamefont {K.}~\bibnamefont {Saito}},\ and\ \bibinfo
  {author} {\bibfnamefont {R.~S.}\ \bibnamefont {Whitney}},\ }\bibfield
  {title} {\bibinfo {title} {Fundamental aspects of steady-state conversion of
  heat to work at the nanoscale},\ }\href
  {https://doi.org/10.1016/j.physrep.2017.05.008} {\bibfield  {journal}
  {\bibinfo  {journal} {Physics Reports}\ }\textbf {\bibinfo {volume} {694}},\
  \bibinfo {pages} {1} (\bibinfo {year} {2017})}\BibitemShut {NoStop}%
\bibitem [{\citenamefont {Brandner}\ \emph {et~al.}(2018)\citenamefont
  {Brandner}, \citenamefont {Hanazato},\ and\ \citenamefont
  {Saito}}]{brandner2018}%
  \BibitemOpen
  \bibfield  {author} {\bibinfo {author} {\bibfnamefont {K.}~\bibnamefont
  {Brandner}}, \bibinfo {author} {\bibfnamefont {T.}~\bibnamefont {Hanazato}},\
  and\ \bibinfo {author} {\bibfnamefont {K.}~\bibnamefont {Saito}},\ }\bibfield
   {title} {\bibinfo {title} {Thermodynamic {{Bounds}} on {{Precision}} in
  {{Ballistic Multiterminal Transport}}},\ }\href
  {https://doi.org/10.1103/PhysRevLett.120.090601} {\bibfield  {journal}
  {\bibinfo  {journal} {Phys. Rev. Lett.}\ }\textbf {\bibinfo {volume} {120}},\
  \bibinfo {pages} {090601} (\bibinfo {year} {2018})}\BibitemShut {NoStop}%
\bibitem [{\citenamefont {Agarwalla}\ \emph {et~al.}(2018)\citenamefont  
  {Agarwalla},\ and\ \citenamefont
  {Segal}}]{agarwalla2018}%
  \BibitemOpen
  \bibfield  {author} {\bibinfo {author} {\bibfnamefont {B.~K.}~\bibnamefont
  {Agarwalla}},\ and\ \bibinfo {author} {\bibfnamefont {D.}~\bibnamefont {Segal}},\ }\bibfield
   {title} {\bibinfo {title} {Assessing the validity of the thermodynamic uncertainty relation in quantum systems},\ }\href
  {https://link.aps.org/doi/10.1103/PhysRevB.98.155438} {\bibfield  {journal}
  {\bibinfo  {journal} {Phys. Rev. B}\ }\textbf {\bibinfo {volume} {98}},\
  \bibinfo {pages} {155438} (\bibinfo {year} {2018})}\BibitemShut {NoStop}%
\bibitem [{\citenamefont {Saryal}\ \emph {et~al.}(2019)\citenamefont {Saryal},
  \citenamefont {Friedman}, \citenamefont {Segal},\ and\ \citenamefont
  {Agarwalla}}]{saryal2019}%
  \BibitemOpen
  \bibfield  {author} {\bibinfo {author} {\bibfnamefont {S.}~\bibnamefont
  {Saryal}}, \bibinfo {author} {\bibfnamefont {H.~M.}\ \bibnamefont
  {Friedman}}, \bibinfo {author} {\bibfnamefont {D.}~\bibnamefont {Segal}},\
  and\ \bibinfo {author} {\bibfnamefont {B.~K.}\ \bibnamefont {Agarwalla}},\
  }\bibfield  {title} {\bibinfo {title} {Thermodynamic uncertainty relation in
  thermal transport},\ }\href {https://doi.org/10.1103/PhysRevE.100.042101}
  {\bibfield  {journal} {\bibinfo  {journal} {Phys. Rev. E}\ }\textbf {\bibinfo
  {volume} {100}},\ \bibinfo {pages} {042101} (\bibinfo {year}
  {2019})}\BibitemShut {NoStop}%
\bibitem [{\citenamefont {Ehrlich}\ and\ \citenamefont
  {Schaller}(2021)}]{ehrlich2021}%
  \BibitemOpen
  \bibfield  {author} {\bibinfo {author} {\bibfnamefont {T.}~\bibnamefont
  {Ehrlich}}\ and\ \bibinfo {author} {\bibfnamefont {G.}~\bibnamefont
  {Schaller}},\ }\bibfield  {title} {\bibinfo {title} {Broadband frequency
  filters with quantum dot chains},\ }\href
  {https://doi.org/10.1103/PhysRevB.104.045424} {\bibfield  {journal} {\bibinfo
   {journal} {Phys. Rev. B}\ }\textbf {\bibinfo {volume} {104}},\ \bibinfo
  {pages} {045424} (\bibinfo {year} {2021})}\BibitemShut {NoStop}%
\bibitem [{\citenamefont {Gerry}\ and\ \citenamefont
  {Segal}(2022)}]{gerry2022}%
  \BibitemOpen
  \bibfield  {author} {\bibinfo {author} {\bibfnamefont {M.}~\bibnamefont
  {Gerry}}\ and\ \bibinfo {author} {\bibfnamefont {D.}~\bibnamefont {Segal}},\
  }\bibfield  {title} {\bibinfo {title} {Absence and recovery of cost-precision
  tradeoff relations in quantum transport},\ }\href
  {https://doi.org/10.1103/PhysRevB.105.155401} {\bibfield  {journal} {\bibinfo
   {journal} {Phys. Rev. B}\ }\textbf {\bibinfo {volume} {105}},\ \bibinfo
  {pages} {155401} (\bibinfo {year} {2022})}\BibitemShut {NoStop}%
\bibitem [{\citenamefont {Timpanaro}\ \emph {et~al.}(2024)\citenamefont
  {Timpanaro}, \citenamefont {Guarnieri},\ and\ \citenamefont
  {Landi}}]{timpanaro2024}%
  \BibitemOpen
  \bibfield  {author} {\bibinfo {author} {\bibfnamefont {A.~M.}\ \bibnamefont
  {Timpanaro}}, \bibinfo {author} {\bibfnamefont {G.}~\bibnamefont
  {Guarnieri}},\ and\ \bibinfo {author} {\bibfnamefont {G.~T.}\ \bibnamefont
  {Landi}},\ }\href@noop {} {\bibinfo {title} {The most accurate quantum
  thermoelectric}} (\bibinfo {year} {2024}),\ \Eprint
  {https://arxiv.org/abs/2106.10205} {arXiv:2106.10205} \BibitemShut {NoStop}%
\bibitem [{\citenamefont {Eriksson}\ \emph {et~al.}(2021)\citenamefont
  {Eriksson}, \citenamefont {Acciai}, \citenamefont {Tesser},\ and\
  \citenamefont {Splettstoesser}}]{eriksson2021}%
  \BibitemOpen
  \bibfield  {author} {\bibinfo {author} {\bibfnamefont {J.}~\bibnamefont
  {Eriksson}}, \bibinfo {author} {\bibfnamefont {M.}~\bibnamefont {Acciai}},
  \bibinfo {author} {\bibfnamefont {L.}~\bibnamefont {Tesser}},\ and\ \bibinfo
  {author} {\bibfnamefont {J.}~\bibnamefont {Splettstoesser}},\ }\bibfield
  {title} {\bibinfo {title} {General {{Bounds}} on {{Electronic Shot Noise}} in
  the {{Absence}} of {{Currents}}},\ }\href
  {https://doi.org/10.1103/PhysRevLett.127.136801} {\bibfield  {journal}
  {\bibinfo  {journal} {Phys. Rev. Lett.}\ }\textbf {\bibinfo {volume} {127}},\
  \bibinfo {pages} {136801} (\bibinfo {year} {2021})}\BibitemShut {NoStop}%
\bibitem [{\citenamefont {Tesser}\ \emph
  {et~al.}(2023{\natexlab{a}})\citenamefont {Tesser}, \citenamefont {Acciai},
  \citenamefont {Sp{\aa}nsl{\"a}tt}, \citenamefont {Monsel},\ and\
  \citenamefont {Splettstoesser}}]{tesser2023}%
  \BibitemOpen
  \bibfield  {author} {\bibinfo {author} {\bibfnamefont {L.}~\bibnamefont
  {Tesser}}, \bibinfo {author} {\bibfnamefont {M.}~\bibnamefont {Acciai}},
  \bibinfo {author} {\bibfnamefont {C.}~\bibnamefont {Sp{\aa}nsl{\"a}tt}},
  \bibinfo {author} {\bibfnamefont {J.}~\bibnamefont {Monsel}},\ and\ \bibinfo
  {author} {\bibfnamefont {J.}~\bibnamefont {Splettstoesser}},\ }\bibfield
  {title} {\bibinfo {title} {Charge, spin, and heat shot noises in the absence
  of average currents: {{Conditions}} on bounds at zero and finite
  frequencies},\ }\href {https://doi.org/10.1103/PhysRevB.107.075409}
  {\bibfield  {journal} {\bibinfo  {journal} {Phys. Rev. B}\ }\textbf {\bibinfo
  {volume} {107}},\ \bibinfo {pages} {075409} (\bibinfo {year}
  {2023}{\natexlab{a}})}\BibitemShut {NoStop}%
\bibitem [{\citenamefont {Tesser}\ and\ \citenamefont
  {Splettstoesser}(2024)}]{tesser2024}%
  \BibitemOpen
  \bibfield  {author} {\bibinfo {author} {\bibfnamefont {L.}~\bibnamefont
  {Tesser}}\ and\ \bibinfo {author} {\bibfnamefont {J.}~\bibnamefont
  {Splettstoesser}},\ }\bibfield  {title} {\bibinfo {title}
  {Out-of-{{Equilibrium Fluctuation-Dissipation Bounds}}},\ }\href
  {https://doi.org/10.1103/PhysRevLett.132.186304} {\bibfield  {journal}
  {\bibinfo  {journal} {Phys. Rev. Lett.}\ }\textbf {\bibinfo {volume} {132}},\
  \bibinfo {pages} {186304} (\bibinfo {year} {2024})}\BibitemShut {NoStop}%
\bibitem [{\citenamefont {Acciai}\ \emph {et~al.}(2024)\citenamefont {Acciai},
  \citenamefont {Tesser}, \citenamefont {Eriksson}, \citenamefont
  {S{\'a}nchez}, \citenamefont {Whitney},\ and\ \citenamefont
  {Splettstoesser}}]{acciai2024}%
  \BibitemOpen
  \bibfield  {author} {\bibinfo {author} {\bibfnamefont {M.}~\bibnamefont
  {Acciai}}, \bibinfo {author} {\bibfnamefont {L.}~\bibnamefont {Tesser}},
  \bibinfo {author} {\bibfnamefont {J.}~\bibnamefont {Eriksson}}, \bibinfo
  {author} {\bibfnamefont {R.}~\bibnamefont {S{\'a}nchez}}, \bibinfo {author}
  {\bibfnamefont {R.~S.}\ \bibnamefont {Whitney}},\ and\ \bibinfo {author}
  {\bibfnamefont {J.}~\bibnamefont {Splettstoesser}},\ }\bibfield  {title}
  {\bibinfo {title} {Constraints between entropy production and its
  fluctuations in nonthermal engines},\ }\href
  {https://doi.org/10.1103/PhysRevB.109.075405} {\bibfield  {journal} {\bibinfo
   {journal} {Phys. Rev. B}\ }\textbf {\bibinfo {volume} {109}},\ \bibinfo
  {pages} {075405} (\bibinfo {year} {2024})}\BibitemShut {NoStop}%
\bibitem [{\citenamefont {Palmqvist}\ \emph {et~al.}(2024)\citenamefont
  {Palmqvist}, \citenamefont {Tesser},\ and\ \citenamefont
  {Splettstoesser}}]{palmqvist2024}%
  \BibitemOpen
  \bibfield  {author} {\bibinfo {author} {\bibfnamefont {D.}~\bibnamefont
  {Palmqvist}}, \bibinfo {author} {\bibfnamefont {L.}~\bibnamefont {Tesser}},\
  and\ \bibinfo {author} {\bibfnamefont {J.}~\bibnamefont {Splettstoesser}},\
  }\href@noop {} {\bibinfo {title} {Kinetic uncertainty relations for quantum
  transport}} (\bibinfo {year} {2024}),\ \Eprint
  {https://arxiv.org/abs/2410.10793} {arXiv:2410.10793} \BibitemShut {NoStop}%
\bibitem [{\citenamefont {Blasi}\ \emph {et~al.}(2025)\citenamefont
  {Blasi}, \citenamefont {Rodríguez}, \citenamefont {Moskalets}, \citenamefont {López},  \ and\ \citenamefont
  {Haack}}]{blasi2025}%
  \BibitemOpen
  \bibfield  {author} {\bibinfo {author} {\bibfnamefont {G.}~\bibnamefont
  {Blasi}}, \bibinfo {author} {\bibfnamefont {R.~R.}~\bibnamefont {Rodríguez}},  \bibinfo {author} {\bibfnamefont {M.}~\bibnamefont {Moskalets}},
  \bibinfo {author} {\bibfnamefont {R.}~\bibnamefont {López}}, \
  and\ \bibinfo {author} {\bibfnamefont {G.}~\bibnamefont {Haack}},\
  }\href@noop {} {\bibinfo {title} {Quantum kinetic uncertainty relations in 
  mesoscopic conductors at strong coupling}} (\bibinfo {year} {2025}),\ \Eprint
  {https://arxiv.org/abs/2505.13200} {arXiv:2505.13200} \BibitemShut {NoStop}%
\bibitem [{\citenamefont {Hasegawa}\ and\ \citenamefont
  {Van~Vu}(2019)}]{hasegawa2019}%
  \BibitemOpen
  \bibfield  {author} {\bibinfo {author} {\bibfnamefont {Y.}~\bibnamefont
  {Hasegawa}}\ and\ \bibinfo {author} {\bibfnamefont {T.}~\bibnamefont
  {Van~Vu}},\ }\bibfield  {title} {\bibinfo {title} {Fluctuation {{Theorem
  Uncertainty Relation}}},\ }\href
  {https://doi.org/10.1103/PhysRevLett.123.110602} {\bibfield  {journal}
  {\bibinfo  {journal} {Phys. Rev. Lett.}\ }\textbf {\bibinfo {volume} {123}},\
  \bibinfo {pages} {110602} (\bibinfo {year} {2019})}\BibitemShut {NoStop}%
\bibitem [{\citenamefont {Timpanaro}\ \emph {et~al.}(2019)\citenamefont
  {Timpanaro}, \citenamefont {Guarnieri}, \citenamefont {Goold},\ and\
  \citenamefont {Landi}}]{timpanaro2019}%
  \BibitemOpen
  \bibfield  {author} {\bibinfo {author} {\bibfnamefont {A.~M.}\ \bibnamefont
  {Timpanaro}}, \bibinfo {author} {\bibfnamefont {G.}~\bibnamefont
  {Guarnieri}}, \bibinfo {author} {\bibfnamefont {J.}~\bibnamefont {Goold}},\
  and\ \bibinfo {author} {\bibfnamefont {G.~T.}\ \bibnamefont {Landi}},\
  }\bibfield  {title} {\bibinfo {title} {Thermodynamic {{Uncertainty
  Relations}} from {{Exchange Fluctuation Theorems}}},\ }\href
  {https://doi.org/10.1103/PhysRevLett.123.090604} {\bibfield  {journal}
  {\bibinfo  {journal} {Phys. Rev. Lett.}\ }\textbf {\bibinfo {volume} {123}},\
  \bibinfo {pages} {090604} (\bibinfo {year} {2019})}\BibitemShut {NoStop}%
\bibitem [{\citenamefont {Potts}\ and\ \citenamefont
  {Samuelsson}(2019)}]{potts2019}%
  \BibitemOpen
  \bibfield  {author} {\bibinfo {author} {\bibfnamefont {P.~P.}\ \bibnamefont
  {Potts}}\ and\ \bibinfo {author} {\bibfnamefont {P.}~\bibnamefont
  {Samuelsson}},\ }\bibfield  {title} {\bibinfo {title} {Thermodynamic
  uncertainty relations including measurement and feedback},\ }\href
  {https://doi.org/10.1103/PhysRevE.100.052137} {\bibfield  {journal} {\bibinfo
   {journal} {Phys. Rev. E}\ }\textbf {\bibinfo {volume} {100}},\ \bibinfo
  {pages} {052137} (\bibinfo {year} {2019})}\BibitemShut {NoStop}%
\bibitem [{\citenamefont {Falasco}\ \emph {et~al.}(2020)\citenamefont
  {Falasco}, \citenamefont {Esposito},\ and\ \citenamefont
  {Delvenne}}]{falasco2020}%
  \BibitemOpen
  \bibfield  {author} {\bibinfo {author} {\bibfnamefont {G.}~\bibnamefont
  {Falasco}}, \bibinfo {author} {\bibfnamefont {M.}~\bibnamefont {Esposito}},\
  and\ \bibinfo {author} {\bibfnamefont {J.-C.}\ \bibnamefont {Delvenne}},\
  }\bibfield  {title} {\bibinfo {title} {Unifying thermodynamic uncertainty
  relations},\ }\href {https://doi.org/10.1088/1367-2630/ab8679} {\bibfield
  {journal} {\bibinfo  {journal} {New J. Phys.}\ }\textbf {\bibinfo {volume}
  {22}},\ \bibinfo {pages} {053046} (\bibinfo {year} {2020})}\BibitemShut
  {NoStop}%
\bibitem [{\citenamefont {B{\"u}ttiker}(1988)}]{buttiker1988}%
  \BibitemOpen
  \bibfield  {author} {\bibinfo {author} {\bibfnamefont {M.}~\bibnamefont
  {B{\"u}ttiker}},\ }\bibfield  {title} {\bibinfo {title} {Coherent and
  sequential tunneling in series barriers},\ }\href
  {https://doi.org/10.1147/rd.321.0063} {\bibfield  {journal} {\bibinfo
  {journal} {IBM J. Res. \& Dev.}\ }\textbf {\bibinfo {volume} {32}},\ \bibinfo
  {pages} {63} (\bibinfo {year} {1988})}\BibitemShut {NoStop}%
\bibitem [{\citenamefont {Brandner}\ and\ \citenamefont
  {Saito}(2025)}]{brandner2025}%
  \BibitemOpen
  \href@noop {} 
  {\bibinfo {title} {See Supplemental Material at [URL will be inserted by publisher] for further details on the derivations of the bounds (2) and (16)}} 
  \BibitemShut {NoStop}%
\bibitem [{\citenamefont {Nenciu}(2007)}]{nenciu2007}%
  \BibitemOpen
  \bibfield  {author} {\bibinfo {author} {\bibfnamefont {G.}~\bibnamefont
  {Nenciu}},\ }\bibfield  {title} {\bibinfo {title} {Independent electron model
  for open quantum systems: {{Landauer-B{\"u}ttiker}} formula and strict
  positivity of the entropy production},\ }\href
  {https://doi.org/10.1063/1.2712418} {\bibfield  {journal} {\bibinfo
  {journal} {J. Math. Phys.}\ }\textbf {\bibinfo {volume} {48}},\ \bibinfo
  {pages} {033302} (\bibinfo {year} {2007})}\BibitemShut {NoStop}%
\bibitem [{\citenamefont {Ji}\ \emph {et~al.}(2003)\citenamefont {Ji},
  \citenamefont {Chung}, \citenamefont {Sprinzak}, \citenamefont {Heiblum},
  \citenamefont {Mahalu},\ and\ \citenamefont {Shtrikman}}]{ji2003}%
  \BibitemOpen
  \bibfield  {author} {\bibinfo {author} {\bibfnamefont {Y.}~\bibnamefont
  {Ji}}, \bibinfo {author} {\bibfnamefont {Y.}~\bibnamefont {Chung}}, \bibinfo
  {author} {\bibfnamefont {D.}~\bibnamefont {Sprinzak}}, \bibinfo {author}
  {\bibfnamefont {M.}~\bibnamefont {Heiblum}}, \bibinfo {author} {\bibfnamefont
  {D.}~\bibnamefont {Mahalu}},\ and\ \bibinfo {author} {\bibfnamefont
  {H.}~\bibnamefont {Shtrikman}},\ }\bibfield  {title} {\bibinfo {title} {An
  electronic {{Mach}}--{{Zehnder}} interferometer},\ }\href
  {https://doi.org/10.1038/nature01503} {\bibfield  {journal} {\bibinfo
  {journal} {Nature}\ }\textbf {\bibinfo {volume} {422}},\ \bibinfo {pages}
  {415} (\bibinfo {year} {2003})}\BibitemShut {NoStop}%
\bibitem [{\citenamefont {Granger}\ \emph {et~al.}(2009)\citenamefont
  {Granger}, \citenamefont {Eisenstein},\ and\ \citenamefont
  {Reno}}]{granger2009}%
  \BibitemOpen
  \bibfield  {author} {\bibinfo {author} {\bibfnamefont {G.}~\bibnamefont
  {Granger}}, \bibinfo {author} {\bibfnamefont {J.~P.}\ \bibnamefont
  {Eisenstein}},\ and\ \bibinfo {author} {\bibfnamefont {J.~L.}\ \bibnamefont
  {Reno}},\ }\bibfield  {title} {\bibinfo {title} {Observation of {{Chiral Heat
  Transport}} in the {{Quantum Hall Regime}}},\ }\href
  {https://doi.org/10.1103/PhysRevLett.102.086803} {\bibfield  {journal}
  {\bibinfo  {journal} {Phys. Rev. Lett.}\ }\textbf {\bibinfo {volume} {102}},\
  \bibinfo {pages} {086803} (\bibinfo {year} {2009})}\BibitemShut {NoStop}%
\bibitem [{\citenamefont {Nam}\ \emph {et~al.}(2013)\citenamefont {Nam},
  \citenamefont {Hwang},\ and\ \citenamefont {Lee}}]{nam2013}%
  \BibitemOpen
  \bibfield  {author} {\bibinfo {author} {\bibfnamefont {S.-G.}\ \bibnamefont
  {Nam}}, \bibinfo {author} {\bibfnamefont {E.~H.}\ \bibnamefont {Hwang}},\
  and\ \bibinfo {author} {\bibfnamefont {H.-J.}\ \bibnamefont {Lee}},\
  }\bibfield  {title} {\bibinfo {title} {Thermoelectric {{Detection}} of
  {{Chiral Heat Transport}} in {{Graphene}} in the {{Quantum Hall Regime}}},\
  }\href {https://doi.org/10.1103/PhysRevLett.110.226801} {\bibfield  {journal}
  {\bibinfo  {journal} {Phys. Rev. Lett.}\ }\textbf {\bibinfo {volume} {110}},\
  \bibinfo {pages} {226801} (\bibinfo {year} {2013})}\BibitemShut {NoStop}%
\bibitem [{\citenamefont {Jezouin}\ \emph {et~al.}(2013)\citenamefont
  {Jezouin}, \citenamefont {Parmentier}, \citenamefont {Anthore}, \citenamefont
  {Gennser}, \citenamefont {Cavanna}, \citenamefont {Jin},\ and\ \citenamefont
  {Pierre}}]{jezouin2013}%
  \BibitemOpen
  \bibfield  {author} {\bibinfo {author} {\bibfnamefont {S.}~\bibnamefont
  {Jezouin}}, \bibinfo {author} {\bibfnamefont {F.~D.}\ \bibnamefont
  {Parmentier}}, \bibinfo {author} {\bibfnamefont {A.}~\bibnamefont {Anthore}},
  \bibinfo {author} {\bibfnamefont {U.}~\bibnamefont {Gennser}}, \bibinfo
  {author} {\bibfnamefont {A.}~\bibnamefont {Cavanna}}, \bibinfo {author}
  {\bibfnamefont {Y.}~\bibnamefont {Jin}},\ and\ \bibinfo {author}
  {\bibfnamefont {F.}~\bibnamefont {Pierre}},\ }\bibfield  {title} {\bibinfo
  {title} {Quantum {{Limit}} of {{Heat Flow Across}} a {{Single Electronic
  Channel}}},\ }\href {https://doi.org/10.1126/science.1241912} {\bibfield
  {journal} {\bibinfo  {journal} {Science}\ }\textbf {\bibinfo {volume}
  {342}},\ \bibinfo {pages} {601} (\bibinfo {year} {2013})}\BibitemShut
  {NoStop}%
\bibitem [{\citenamefont {B{\"a}uerle}\ \emph {et~al.}(2018)\citenamefont
  {B{\"a}uerle}, \citenamefont {Christian~Glattli}, \citenamefont {Meunier},
  \citenamefont {Portier}, \citenamefont {Roche}, \citenamefont {Roulleau},
  \citenamefont {Takada},\ and\ \citenamefont {Waintal}}]{bauerle2018}%
  \BibitemOpen
  \bibfield  {author} {\bibinfo {author} {\bibfnamefont {C.}~\bibnamefont
  {B{\"a}uerle}}, \bibinfo {author} {\bibfnamefont {D.}~\bibnamefont
  {Christian~Glattli}}, \bibinfo {author} {\bibfnamefont {T.}~\bibnamefont
  {Meunier}}, \bibinfo {author} {\bibfnamefont {F.}~\bibnamefont {Portier}},
  \bibinfo {author} {\bibfnamefont {P.}~\bibnamefont {Roche}}, \bibinfo
  {author} {\bibfnamefont {P.}~\bibnamefont {Roulleau}}, \bibinfo {author}
  {\bibfnamefont {S.}~\bibnamefont {Takada}},\ and\ \bibinfo {author}
  {\bibfnamefont {X.}~\bibnamefont {Waintal}},\ }\bibfield  {title} {\bibinfo
  {title} {Coherent control of single electrons: A review of current
  progress},\ }\href {https://doi.org/10.1088/1361-6633/aaa98a} {\bibfield
  {journal} {\bibinfo  {journal} {Rep. Prog. Phys.}\ }\textbf {\bibinfo
  {volume} {81}},\ \bibinfo {pages} {056503} (\bibinfo {year}
  {2018})}\BibitemShut {NoStop}%
\bibitem [{\citenamefont {Sivre}\ \emph {et~al.}(2019)\citenamefont {Sivre},
  \citenamefont {Duprez}, \citenamefont {Anthore}, \citenamefont {Aassime},
  \citenamefont {Parmentier}, \citenamefont {Cavanna}, \citenamefont {Ouerghi},
  \citenamefont {Gennser},\ and\ \citenamefont {Pierre}}]{sivre2019}%
  \BibitemOpen
  \bibfield  {author} {\bibinfo {author} {\bibfnamefont {E.}~\bibnamefont
  {Sivre}}, \bibinfo {author} {\bibfnamefont {H.}~\bibnamefont {Duprez}},
  \bibinfo {author} {\bibfnamefont {A.}~\bibnamefont {Anthore}}, \bibinfo
  {author} {\bibfnamefont {A.}~\bibnamefont {Aassime}}, \bibinfo {author}
  {\bibfnamefont {F.~D.}\ \bibnamefont {Parmentier}}, \bibinfo {author}
  {\bibfnamefont {A.}~\bibnamefont {Cavanna}}, \bibinfo {author} {\bibfnamefont
  {A.}~\bibnamefont {Ouerghi}}, \bibinfo {author} {\bibfnamefont
  {U.}~\bibnamefont {Gennser}},\ and\ \bibinfo {author} {\bibfnamefont
  {F.}~\bibnamefont {Pierre}},\ }\bibfield  {title} {\bibinfo {title}
  {Electronic heat flow and thermal shot noise in quantum circuits},\ }\href
  {https://doi.org/10.1038/s41467-019-13566-8} {\bibfield  {journal} {\bibinfo
  {journal} {Nat. Commun.}\ }\textbf {\bibinfo {volume} {10}},\ \bibinfo {pages}
  {5638} (\bibinfo {year} {2019})}\BibitemShut {NoStop}%
\bibitem [{\citenamefont {Whitney}(2014)}]{whitney2014}%
  \BibitemOpen
  \bibfield  {author} {\bibinfo {author} {\bibfnamefont {R.~S.}\ \bibnamefont
  {Whitney}},\ }\bibfield  {title} {\bibinfo {title} {Most {{Efficient Quantum
  Thermoelectric}} at {{Finite Power Output}}},\ }\href
  {https://doi.org/10.1103/PhysRevLett.112.130601} {\bibfield  {journal}
  {\bibinfo  {journal} {Phys. Rev. Lett.}\ }\textbf {\bibinfo {volume} {112}},\
  \bibinfo {pages} {130601} (\bibinfo {year} {2014})}\BibitemShut {NoStop}%
\bibitem [{\citenamefont {Whitney}(2015)}]{whitney2015}%
  \BibitemOpen
  \bibfield  {author} {\bibinfo {author} {\bibfnamefont {R.~S.}\ \bibnamefont
  {Whitney}},\ }\bibfield  {title} {\bibinfo {title} {Finding the quantum
  thermoelectric with maximal efficiency and minimal entropy production at
  given power output},\ }\href {https://doi.org/10.1103/PhysRevB.91.115425}
  {\bibfield  {journal} {\bibinfo  {journal} {Phys. Rev. B}\ }\textbf {\bibinfo
  {volume} {91}},\ \bibinfo {pages} {115425} (\bibinfo {year}
  {2015})}\BibitemShut {NoStop}%
\bibitem [{\citenamefont {Tesser}\ \emph
  {et~al.}(2023{\natexlab{b}})\citenamefont {Tesser}, \citenamefont {Whitney},\
  and\ \citenamefont {Splettstoesser}}]{tesser2023a}%
  \BibitemOpen
  \bibfield  {author} {\bibinfo {author} {\bibfnamefont {L.}~\bibnamefont
  {Tesser}}, \bibinfo {author} {\bibfnamefont {R.~S.}\ \bibnamefont
  {Whitney}},\ and\ \bibinfo {author} {\bibfnamefont {J.}~\bibnamefont
  {Splettstoesser}},\ }\bibfield  {title} {\bibinfo {title} {Thermodynamic
  {{Performance}} of {{Hot-Carrier Solar Cells}}: {{A Quantum Transport
  Model}}},\ }\href {https://doi.org/10.1103/PhysRevApplied.19.044038}
  {\bibfield  {journal} {\bibinfo  {journal} {Phys. Rev. Applied}\ }\textbf
  {\bibinfo {volume} {19}},\ \bibinfo {pages} {044038} (\bibinfo {year}
  {2023}{\natexlab{b}})}\BibitemShut {NoStop}%
\bibitem [{\citenamefont {Mishra}\ and\ \citenamefont
  {Benjamin}(2024)}]{mishra2024}%
  \BibitemOpen
  \bibfield  {author} {\bibinfo {author} {\bibfnamefont {S.}~\bibnamefont
  {Mishra}}\ and\ \bibinfo {author} {\bibfnamefont {C.}~\bibnamefont
  {Benjamin}},\ }\bibfield  {title} {\bibinfo {title} {Reaching the {{Van}} den
  {{Broeck}} limit in linear response and the {{Whitney}} limit in nonlinear
  response in edge mode quantum thermoelectrics and refrigeration},\ }\href
  {https://doi.org/10.1103/PhysRevB.110.235430} {\bibfield  {journal} {\bibinfo
   {journal} {Phys. Rev. B}\ }\textbf {\bibinfo {volume} {110}},\ \bibinfo
  {pages} {235430} (\bibinfo {year} {2024})}\BibitemShut {NoStop}%
\bibitem [{\citenamefont {Volk}\ \emph {et~al.}(2019)\citenamefont {Volk},
  \citenamefont {Zwerver}, \citenamefont {Mukhopadhyay}, \citenamefont
  {Eendebak}, \citenamefont {Van~Diepen}, \citenamefont {Dehollain},
  \citenamefont {Hensgens}, \citenamefont {Fujita}, \citenamefont {Reichl},
  \citenamefont {Wegscheider},\ and\ \citenamefont {Vandersypen}}]{volk2019}%
  \BibitemOpen
  \bibfield  {author} {\bibinfo {author} {\bibfnamefont {C.}~\bibnamefont
  {Volk}}, \bibinfo {author} {\bibfnamefont {A.~M.~J.}\ \bibnamefont
  {Zwerver}}, \bibinfo {author} {\bibfnamefont {U.}~\bibnamefont
  {Mukhopadhyay}}, \bibinfo {author} {\bibfnamefont {P.~T.}\ \bibnamefont
  {Eendebak}}, \bibinfo {author} {\bibfnamefont {C.~J.}\ \bibnamefont
  {Van~Diepen}}, \bibinfo {author} {\bibfnamefont {J.~P.}\ \bibnamefont
  {Dehollain}}, \bibinfo {author} {\bibfnamefont {T.}~\bibnamefont {Hensgens}},
  \bibinfo {author} {\bibfnamefont {T.}~\bibnamefont {Fujita}}, \bibinfo
  {author} {\bibfnamefont {C.}~\bibnamefont {Reichl}}, \bibinfo {author}
  {\bibfnamefont {W.}~\bibnamefont {Wegscheider}},\ and\ \bibinfo {author}
  {\bibfnamefont {L.~M.~K.}\ \bibnamefont {Vandersypen}},\ }\bibfield  {title}
  {\bibinfo {title} {Loading a quantum-dot based ``{{Qubyte}}'' register},\
  }\href {https://doi.org/10.1038/s41534-019-0146-y} {\bibfield  {journal}
  {\bibinfo  {journal} {npj Quantum Inf.}\ }\textbf {\bibinfo {volume} {5}},\
  \bibinfo {pages} {29} (\bibinfo {year} {2019})}\BibitemShut {NoStop}%
\bibitem [{\citenamefont {Mills}\ \emph
  {et~al.}(2019{\natexlab{a}})\citenamefont {Mills}, \citenamefont {Zajac},
  \citenamefont {Gullans}, \citenamefont {Schupp}, \citenamefont {Hazard},\
  and\ \citenamefont {Petta}}]{mills2019}%
  \BibitemOpen
  \bibfield  {author} {\bibinfo {author} {\bibfnamefont {A.~R.}\ \bibnamefont
  {Mills}}, \bibinfo {author} {\bibfnamefont {D.~M.}\ \bibnamefont {Zajac}},
  \bibinfo {author} {\bibfnamefont {M.~J.}\ \bibnamefont {Gullans}}, \bibinfo
  {author} {\bibfnamefont {F.~J.}\ \bibnamefont {Schupp}}, \bibinfo {author}
  {\bibfnamefont {T.~M.}\ \bibnamefont {Hazard}},\ and\ \bibinfo {author}
  {\bibfnamefont {J.~R.}\ \bibnamefont {Petta}},\ }\bibfield  {title} {\bibinfo
  {title} {Shuttling a single charge across a one-dimensional array of silicon
  quantum dots},\ }\href {https://doi.org/10.1038/s41467-019-08970-z}
  {\bibfield  {journal} {\bibinfo  {journal} {Nat. Commun.}\ }\textbf {\bibinfo
  {volume} {10}},\ \bibinfo {pages} {1063} (\bibinfo {year}
  {2019}{\natexlab{a}})}\BibitemShut {NoStop}%
\bibitem [{\citenamefont {Mills}\ \emph
  {et~al.}(2019{\natexlab{b}})\citenamefont {Mills}, \citenamefont {Feldman},
  \citenamefont {Monical}, \citenamefont {Lewis}, \citenamefont {Larson},
  \citenamefont {Mounce},\ and\ \citenamefont {Petta}}]{mills2019a}%
  \BibitemOpen
  \bibfield  {author} {\bibinfo {author} {\bibfnamefont {A.~R.}\ \bibnamefont
  {Mills}}, \bibinfo {author} {\bibfnamefont {M.~M.}\ \bibnamefont {Feldman}},
  \bibinfo {author} {\bibfnamefont {C.}~\bibnamefont {Monical}}, \bibinfo
  {author} {\bibfnamefont {P.~J.}\ \bibnamefont {Lewis}}, \bibinfo {author}
  {\bibfnamefont {K.~W.}\ \bibnamefont {Larson}}, \bibinfo {author}
  {\bibfnamefont {A.~M.}\ \bibnamefont {Mounce}},\ and\ \bibinfo {author}
  {\bibfnamefont {J.~R.}\ \bibnamefont {Petta}},\ }\bibfield  {title} {\bibinfo
  {title} {Computer-automated tuning procedures for semiconductor quantum dot
  arrays},\ }\href {https://doi.org/10.1063/1.5121444} {\bibfield  {journal}
  {\bibinfo  {journal} {Appl. Phys. Lett.}\ }\textbf {\bibinfo {volume}
  {115}},\ \bibinfo {pages} {113501} (\bibinfo {year}
  {2019}{\natexlab{b}})}\BibitemShut {NoStop}%
\bibitem [{\citenamefont {Qiao}\ \emph {et~al.}(2020)\citenamefont {Qiao},
  \citenamefont {Kandel}, \citenamefont {Deng}, \citenamefont {Fallahi},
  \citenamefont {Gardner}, \citenamefont {Manfra}, \citenamefont {Barnes},\
  and\ \citenamefont {Nichol}}]{qiao2020}%
  \BibitemOpen
  \bibfield  {author} {\bibinfo {author} {\bibfnamefont {H.}~\bibnamefont
  {Qiao}}, \bibinfo {author} {\bibfnamefont {Y.~P.}\ \bibnamefont {Kandel}},
  \bibinfo {author} {\bibfnamefont {K.}~\bibnamefont {Deng}}, \bibinfo {author}
  {\bibfnamefont {S.}~\bibnamefont {Fallahi}}, \bibinfo {author} {\bibfnamefont
  {G.~C.}\ \bibnamefont {Gardner}}, \bibinfo {author} {\bibfnamefont {M.~J.}\
  \bibnamefont {Manfra}}, \bibinfo {author} {\bibfnamefont {E.}~\bibnamefont
  {Barnes}},\ and\ \bibinfo {author} {\bibfnamefont {J.~M.}\ \bibnamefont
  {Nichol}},\ }\bibfield  {title} {\bibinfo {title} {Coherent {{Multispin
  Exchange Coupling}} in a {{Quantum-Dot Spin Chain}}},\ }\href
  {https://doi.org/10.1103/PhysRevX.10.031006} {\bibfield  {journal} {\bibinfo
  {journal} {Phys. Rev. X}\ }\textbf {\bibinfo {volume} {10}},\ \bibinfo
  {pages} {031006} (\bibinfo {year} {2020})}\BibitemShut {NoStop}%
\bibitem [{\citenamefont {Hsiao}\ \emph {et~al.}(2020)\citenamefont {Hsiao},
  \citenamefont {Van~Diepen}, \citenamefont {Mukhopadhyay}, \citenamefont
  {Reichl}, \citenamefont {Wegscheider},\ and\ \citenamefont
  {Vandersypen}}]{hsiao2020}%
  \BibitemOpen
  \bibfield  {author} {\bibinfo {author} {\bibfnamefont {T.-K.}\ \bibnamefont
  {Hsiao}}, \bibinfo {author} {\bibfnamefont {C.}~\bibnamefont {Van~Diepen}},
  \bibinfo {author} {\bibfnamefont {U.}~\bibnamefont {Mukhopadhyay}}, \bibinfo
  {author} {\bibfnamefont {C.}~\bibnamefont {Reichl}}, \bibinfo {author}
  {\bibfnamefont {W.}~\bibnamefont {Wegscheider}},\ and\ \bibinfo {author}
  {\bibfnamefont {L.}~\bibnamefont {Vandersypen}},\ }\bibfield  {title}
  {\bibinfo {title} {Efficient {{Orthogonal Control}} of {{Tunnel Couplings}}
  in a {{Quantum Dot Array}}},\ }\href
  {https://doi.org/10.1103/PhysRevApplied.13.054018} {\bibfield  {journal}
  {\bibinfo  {journal} {Phys. Rev. Applied}\ }\textbf {\bibinfo {volume}
  {13}},\ \bibinfo {pages} {054018} (\bibinfo {year} {2020})}\BibitemShut
  {NoStop}%
\bibitem [{\citenamefont {Kandel}\ \emph {et~al.}(2021)\citenamefont {Kandel},
  \citenamefont {Qiao}, \citenamefont {Fallahi}, \citenamefont {Gardner},
  \citenamefont {Manfra},\ and\ \citenamefont {Nichol}}]{kandel2021}%
  \BibitemOpen
  \bibfield  {author} {\bibinfo {author} {\bibfnamefont {Y.~P.}\ \bibnamefont
  {Kandel}}, \bibinfo {author} {\bibfnamefont {H.}~\bibnamefont {Qiao}},
  \bibinfo {author} {\bibfnamefont {S.}~\bibnamefont {Fallahi}}, \bibinfo
  {author} {\bibfnamefont {G.~C.}\ \bibnamefont {Gardner}}, \bibinfo {author}
  {\bibfnamefont {M.~J.}\ \bibnamefont {Manfra}},\ and\ \bibinfo {author}
  {\bibfnamefont {J.~M.}\ \bibnamefont {Nichol}},\ }\bibfield  {title}
  {\bibinfo {title} {Adiabatic quantum state transfer in a semiconductor
  quantum-dot spin chain},\ }\href {https://doi.org/10.1038/s41467-021-22416-5}
  {\bibfield  {journal} {\bibinfo  {journal} {Nat. Commun.}\ }\textbf {\bibinfo
  {volume} {12}},\ \bibinfo {pages} {2156} (\bibinfo {year}
  {2021})}\BibitemShut {NoStop}%
\bibitem [{\citenamefont {Van~Diepen}\ \emph {et~al.}(2021)\citenamefont
  {Van~Diepen}, \citenamefont {Hsiao}, \citenamefont {Mukhopadhyay},
  \citenamefont {Reichl}, \citenamefont {Wegscheider},\ and\ \citenamefont
  {Vandersypen}}]{vandiepen2021}%
  \BibitemOpen
  \bibfield  {author} {\bibinfo {author} {\bibfnamefont {C.~J.}\ \bibnamefont
  {Van~Diepen}}, \bibinfo {author} {\bibfnamefont {T.-K.}\ \bibnamefont
  {Hsiao}}, \bibinfo {author} {\bibfnamefont {U.}~\bibnamefont {Mukhopadhyay}},
  \bibinfo {author} {\bibfnamefont {C.}~\bibnamefont {Reichl}}, \bibinfo
  {author} {\bibfnamefont {W.}~\bibnamefont {Wegscheider}},\ and\ \bibinfo
  {author} {\bibfnamefont {L.~M.~K.}\ \bibnamefont {Vandersypen}},\ }\bibfield
  {title} {\bibinfo {title} {Quantum {{Simulation}} of {{Antiferromagnetic
  Heisenberg Chain}} with {{Gate-Defined Quantum Dots}}},\ }\href
  {https://doi.org/10.1103/PhysRevX.11.041025} {\bibfield  {journal} {\bibinfo
  {journal} {Phys. Rev. X}\ }\textbf {\bibinfo {volume} {11}},\ \bibinfo
  {pages} {041025} (\bibinfo {year} {2021})}\BibitemShut {NoStop}%
\bibitem [{Note1()}]{Note1}%
  \BibitemOpen
  \bibinfo {note} {We have verified this result using computer algebra for
  $N=1,\protect \dots ,10$ and numerically for $N=11, \protect \dots ,
  100$.}\BibitemShut {Stop}%
\bibitem [{\citenamefont {Brandner}(2020)}]{brandner2020}%
  \BibitemOpen
  \bibfield  {author} {\bibinfo {author} {\bibfnamefont {K.}~\bibnamefont
  {Brandner}},\ }\bibfield  {title} {\bibinfo {title} {Coherent {{Transport}}
  in {{Periodically Driven Mesoscopic Conductors}}: {{From Scattering
  Amplitudes}} to {{Quantum Thermodynamics}}},\ }\href
  {https://doi.org/10.1515/zna-2020-0056} {\bibfield  {journal} {\bibinfo
  {journal} {Z. Naturforsch.}\ }\textbf {\bibinfo {volume} {75}},\ \bibinfo
  {pages} {483} (\bibinfo {year} {2020})}\BibitemShut {NoStop}%
\bibitem [{\citenamefont {Ohnmacht}\ \emph {et~al.}(2024)\citenamefont
  {Ohnmacht}, \citenamefont {Cuevas}, \citenamefont {Belzig}, \citenamefont
  {L{\'o}pez}, \citenamefont {Lim},\ and\ \citenamefont {Kim}}]{ohnmacht2024}%
  \BibitemOpen
  \bibfield  {author} {\bibinfo {author} {\bibfnamefont {D.~C.}\ \bibnamefont
  {Ohnmacht}}, \bibinfo {author} {\bibfnamefont {J.~C.}\ \bibnamefont
  {Cuevas}}, \bibinfo {author} {\bibfnamefont {W.}~\bibnamefont {Belzig}},
  \bibinfo {author} {\bibfnamefont {R.}~\bibnamefont {L{\'o}pez}}, \bibinfo
  {author} {\bibfnamefont {J.~S.}\ \bibnamefont {Lim}},\ and\ \bibinfo {author}
  {\bibfnamefont {K.~W.}\ \bibnamefont {Kim}},\ }\href@noop {} {\bibinfo
  {title} {Thermodynamic uncertainty relations in superconducting junctions}}
  (\bibinfo {year} {2024}),\ \Eprint {https://arxiv.org/abs/2408.01281}
  {arXiv:2408.01281} \BibitemShut {NoStop}%
\end{thebibliography}
\end{document}



\title{Thermodynamic Uncertainty Relations for Coherent Transport:\\ Supplemental Material}
\author{Kay Brandner\textsuperscript{1,2}, Keiji Saito\textsuperscript{3}}
\affiliation{
\textsuperscript{1}School of Physics and Astronomy, University of Nottingham, Nottingham NG7 2RD, United Kingdom\\
\textsuperscript{2}Centre for the Mathematics and Theoretical Physics of Quantum Non-Equilibrium Systems, University of Nottingham, Nottingham NG7 2RD, United Kingdom\\
\textsuperscript{3}Department of Physics, Kyoto University, Kyoto 606-8502, Japan
}





\maketitle

\vbadness=10000
\hbadness=10000

\renewcommand{\theequation}{S\arabic{equation}}

\makeatletter
\newcommand{\manuallabel}[2]{\def\@currentlabel{#2}\label{#1}}
\makeatother

\newcommand{\eb}{Eq.}

\newcommand{\kb}{k_\mathrm{B}}

\newcommand{\Eint}{\int_0^\infty \! dE \;}

\newcommand{\ket}[1]{|#1\rangle}
\newcommand{\bra}[1]{\langle #1|}

\manuallabel{MT_QTUR}{2}
\manuallabel{MT_QTURB}{16}
\manuallabel{MT_HeatCurr}{4}
\manuallabel{MT_EntProd}{9}
\manuallabel{MT_SymEntBnd}{10}

\manuallabel{MT_ARKinEq}{6}
\manuallabel{MT_ExHamiltonian}{22}
\manuallabel{MT_ARForceOp}{15}
\manuallabel{MT_CohFroceOp}{18}
\manuallabel{MT_ExProt}{23}
\manuallabel{MT_Speed}{10}
\manuallabel{MT_SInOut}{1}
\manuallabel{MT_CohTSym}{17}
\manuallabel{MT_CohDecompAdRespCoeff}{19}

\vbadness=10000
\hbadness=10000

\allowdisplaybreaks

In this Supplemental Material, we provide further details on the derivations of the thermodynamic uncertainty relations \eqref{MT_QTUR} and \eqref{MT_QTURB} stated in the main text.  
We begin by observing that, upon inserting Eq.~\eqref{MT_HeatCurr} into Eq.~\eqref{MT_EntProd}, the total rate of entropy production becomes
\begin{align}
\label{eq:SMEntProd}
\sigma 	& = \frac{1}{h}\Eint\sum_{\alpha\beta}\mathcal{T}^{\alpha\beta}_E(f^\beta_E - f^\alpha_E)
			(E-\mu_\alpha)/T_\alpha\\
		& = \frac{\kb}{h}\Eint\sum_{\alpha\beta}\mathcal{T}^{\alpha\beta}_E(f^\beta_E - f^\alpha_E)
			\ln [(1-f^\alpha_E)/f^\alpha_E]. \nonumber
\end{align}
If $\mathcal{T}^{\alpha\beta}_E = \mathcal{T}^{\beta\alpha}_E$, the second line can  be rewritten in the form 
\begin{align}
\sigma 	& = \frac{\kb}{2h}\Eint\sum_{\alpha\beta}\mathcal{T}^{\alpha\beta}_E(f^\alpha_E - f^\beta_E)
			\ln[g^{\alpha\beta}_E / g^{\beta \alpha}_E]\\
		& = \frac{\kb}{2h}\Eint\sum_{\gamma, \beta\neq\alpha}
			\mathcal{T}^{\gamma\beta}_E(g^{\gamma\beta}_E - g^{\beta\gamma}_E)
			\ln[g^{\gamma\beta}_E / g^{\beta \gamma}_E]
		  + \frac{\kb}{h}\Eint\sum_{\beta\neq\alpha}\mathcal{T}^{\alpha\beta}_E(f^\alpha_E - f^\beta_E)
			\ln[g^{\alpha\beta}_E / g^{\beta \alpha}_E],\nonumber
\end{align}
where $g^{\alpha\beta}_E = f^\alpha_E (1-f^\beta_E)> 0$ for any finite $E$. 
Since $(x-y)\ln[x/y]> 0$ for any $x,y> 0$, this result implies Eq.~\eqref{MT_SymEntBnd}, from which Eq.~\eqref{MT_QTUR} follows along the steps outlined in the main text. 
If $\mathcal{T}^{\alpha\beta}_E \neq \mathcal{T}^{\beta\alpha}_E$, we can still use the sum rules $\sum_\beta\mathcal{T}^{\alpha\beta}_E = \sum_\beta\mathcal{T}^{\beta\alpha}_E$ to rewrite the expression \eqref{eq:SMEntProd} as a sum of non-negative terms, which is given by 
\begin{equation}\label{eq:SMEntProdB}
	\sigma  = \frac{\kb}{h}\Eint\sum_{\alpha\beta}\mathcal{T}^{\alpha\beta}_E\bigl\{
				\rho[f^\alpha_E]-\rho[f^\beta_E]-\rho'[f^\alpha_E](f^\alpha_E-f^\beta_E)\bigr\}.
\end{equation}
Here, $\rho[x] = -x\ln[x] -(1-x)\ln[1-x]$ is a concave function over the open interval $(0,1)$ \cite{nenciu2007}.
That is, we have $\rho[x]-\rho[y]\geq \rho'[x](x-y)$ for any $x,y\in (0,1)$. 
Since the Fermi functions take values between $0$ and $1$ for any finite $E$, Eq.~\eqref{eq:SMEntProdB} therefore implies 
\begin{align}
\label{eq:SMEntBndB}
	\sigma  & \geq \frac{\kb}{h}\Eint\sum_{\beta\neq\alpha}\mathcal{T}^{\alpha\beta}_E\bigl\{
				\rho[f^\alpha_E]-\rho[f^\beta_E]-\rho'[f^\alpha_E](f^\alpha_E-f^\beta_E)\bigr\}\\
			& = \frac{\kb}{h}\Eint \sum_{\beta\neq\alpha}\mathcal{T}^{\alpha\beta}_E
	(g^{\alpha\alpha}_E + g^{\beta\beta}_E)\Phi[Y^{\alpha\beta}_E]\psi[f^\alpha_E,f^\beta_E] \nonumber
\end{align}
for any $\alpha$, where we have introduced the functions  
\begin{equation}
	\Phi[x] = x\cdot\text{arsinh}[x] = x\cdot \ln [x+[1+x^2]^\frac{1}{2}]
	\quad\text{and}\quad
		\psi[x,y] = \frac{\rho[x]-\rho[y]-\rho'[x](x-y)}{\mathrm{arsinh}[(x-y)/(x+y-x^2-y^2)](x-y)},
\end{equation}
and the variables $	Y^{\alpha\beta}_E = (f^\alpha_E - f^\beta_E)/(g^{\alpha\alpha}_E + g^{\beta\beta}_E)$. 
Numerical minimization shows that $\psi[x,y]\geq \psi_0 \simeq 0.85246 > 17/20$ for $(x,y)\in [0,1]\times [0,1]$.
Thus, since $\Phi[x]\geq 0$ for any real $x$, the inequality \eqref{eq:SMEntBndB} remains valid after replacing $\psi$ with $\psi_0$. 
After this replacement, we can again proceed along the lines described in the main text.  
That is, by using that $\Phi$ is a convex function and that $S^\text{th}_\alpha\leq S_\alpha$, we arrive at Eq.~\eqref{MT_QTURB}.